\documentclass[prr,twocolumn,aps,showpacs,superscriptaddress,longbibliography]{revtex4-1} 

\usepackage{amsmath}
\usepackage{amssymb}
\usepackage{amsmath}
\usepackage{txfonts}
\usepackage{graphicx} 
\usepackage{epsfig}
\usepackage{float} 
\usepackage[T1]{fontenc}
\usepackage{times}
\usepackage{color} 
\usepackage{cases}
\usepackage{xspace}
\usepackage{amssymb,amsmath}
\usepackage{amsbsy}
\usepackage{braket}
\usepackage{tabularx}
\usepackage{mathtools}          

\usepackage[utf8]{inputenc}

\DeclareMathOperator{\sech}{sech}

\usepackage{ifpdf}
\ifpdf
\usepackage{epstopdf}
\fi

\usepackage[unicode]{hyperref}
\hypersetup{
	pdftitle={spinnorBEC},
	pdfauthor={Yu},
	pdfsubject={Vortexlines},
	colorlinks=true,
	linkcolor=blue,
	citecolor=blue,
	filecolor=black,
	urlcolor=blue
}
\def\urlprefix{}
\def\url#1{}

\usepackage{bm}
\usepackage{color}

\newcommand{\be}{\begin{equation}}

\newcommand{\ee}{\end{equation}}

\newcommand{\bea}{\begin{eqnarray}}

\newcommand{\eea}{\end{eqnarray}}

\newcommand{\nn}{\nonumber }

\begin{document}

\title{Dark soliton-like magnetic domain walls in a two-dimensional ferromagnetic superfluid }
\author{Xiaoquan Yu}\email{xqyu@gscaep.ac.cn}
\affiliation{Graduate School of  China Academy of Engineering Physics, Beijing 100193, China}
\affiliation{Department of Physics, Centre for Quantum Science, and Dodd-Walls Centre for Photonic and Quantum Technologies, University of Otago, Dunedin, New Zealand}

\author{P.~B.~Blakie}\email{blair.blakie@otago.ac.nz
}
\affiliation{Department of Physics, Centre for Quantum Science, and Dodd-Walls Centre for Photonic and Quantum Technologies, University of Otago, Dunedin, New Zealand}

\begin{abstract}
We report a stable magnetic domain wall in a uniform ferromagnetic spin-1  condensate,  characterized by the magnetization having a dark soliton profile with nonvanishing superfluid density.  We find exact stationary solutions for a particular ratio of interaction parameters with and without magnetic fields, and develop an accurate analytic solution applicable to the whole ferromagnetic phase.  In the absence of magnetic fields, this domain wall relates 
various distinct solitary excitations in binary condensates through $\textrm{SO}(3)$ spin rotations, which otherwise are unconnected.  Remarkably, studying the dynamics of a quasi-two-dimensional (quasi-2D) system we show that standing wave excitations of the domain wall oscillate without decay, being stable against the snake instability. The domain wall is dynamically unstable to modes that cause the magnetization to grow perpendicularly while leaving the domain wall unchanged. Real time dynamics in the presence of white noise reveals that this ``spin twist'' instability  does not destroy the topological structure of the magnetic domain wall.

\end{abstract}

\maketitle

\section{Introduction}
A domain wall is a nonlinear excitation that  interpolates between two different ground states, playing an important role in both equilibrium and out-of-equilibrium  phase transitions  with discrete symmetry breaking~\cite{bunkov2012topological,manton2004topological,eto2014vortices,bunkov2012topological}. It appears in broad fields of physics, ranging from statistical mechanics~\cite{bunkov2012topological} and quantum field theories~\cite{manton2004topological,eto2014vortices} to cosmology~\cite{vilenkin2000cosmic}.  

Bose-Einstein condensates (BECs) provide a platform to study various topological excitations including vortices, domain walls and solitons. Unlike vortices, 
a wide  class  of domain walls and solitons  are unstable to the so-called snake instability in two-dimensional (2D) systems, when
the size of the system is larger than the
width of domain walls/solitons.  Examples include dark solitons~\cite{kuznetsov1988instability,Shlyapnikov1999,DSdecayExp,PhysRevA.67.023604}, phase domain walls~\cite{Son2002,Kasamatsu2019,Gallemi2019}, magnetic solitons~\cite{MDQu2016,MSexp1,MSexp2}, and nematic domain wall-vortex composites~\cite{shin2019}, in scalar, coherently coupled, binary, and anti-ferromagnetic spin-1 BECs, respectively. An outstanding challenge is thus to obtain stable  2D domain walls/solitons, which would open the door to studying their rich  dynamical properties.

Thanks to the $\textrm{U}(1)$ gauge symmetry and the rotational $\textrm{SO}(3)$ symmetry, a spin-1 ferromagnetic BEC exhibits both superfluid and magnetic order quantified by the superfluid density and the magnetization~\cite{Ho98,OM98,Stampernatrue2006,StamperRMP,KAWAGUCHI12}, respectively. It offers an opportunity to explore  magnetic domain walls (interfaces separating oppositely magnetized regions) absent in scalar, binary and anti-ferromagnetic BECs.  Most work in ferromagnetic spin-1 BECs has focused on  spin  textures and their the nonequilibrium dynamics (e.g.~\cite{zhang2005DW,Higbie2005,Saito2005DW,Saito2007pcvformation,Saito2007pcvformation,zhang2007,Vengalattore2008a,Kawaguchi2010a,Williamson2016a,Prufer2018a}). The domain wall physics remains largely unexplored and very little is known about their structures, stability in high dimensions and potential connections to vector solitons~\cite{nistazakis2008bright,Busch2001,ThreeComponentSoliton2018}.

\begin{figure}[htp] 
	\centering 
	\includegraphics[width=0.47\textwidth]{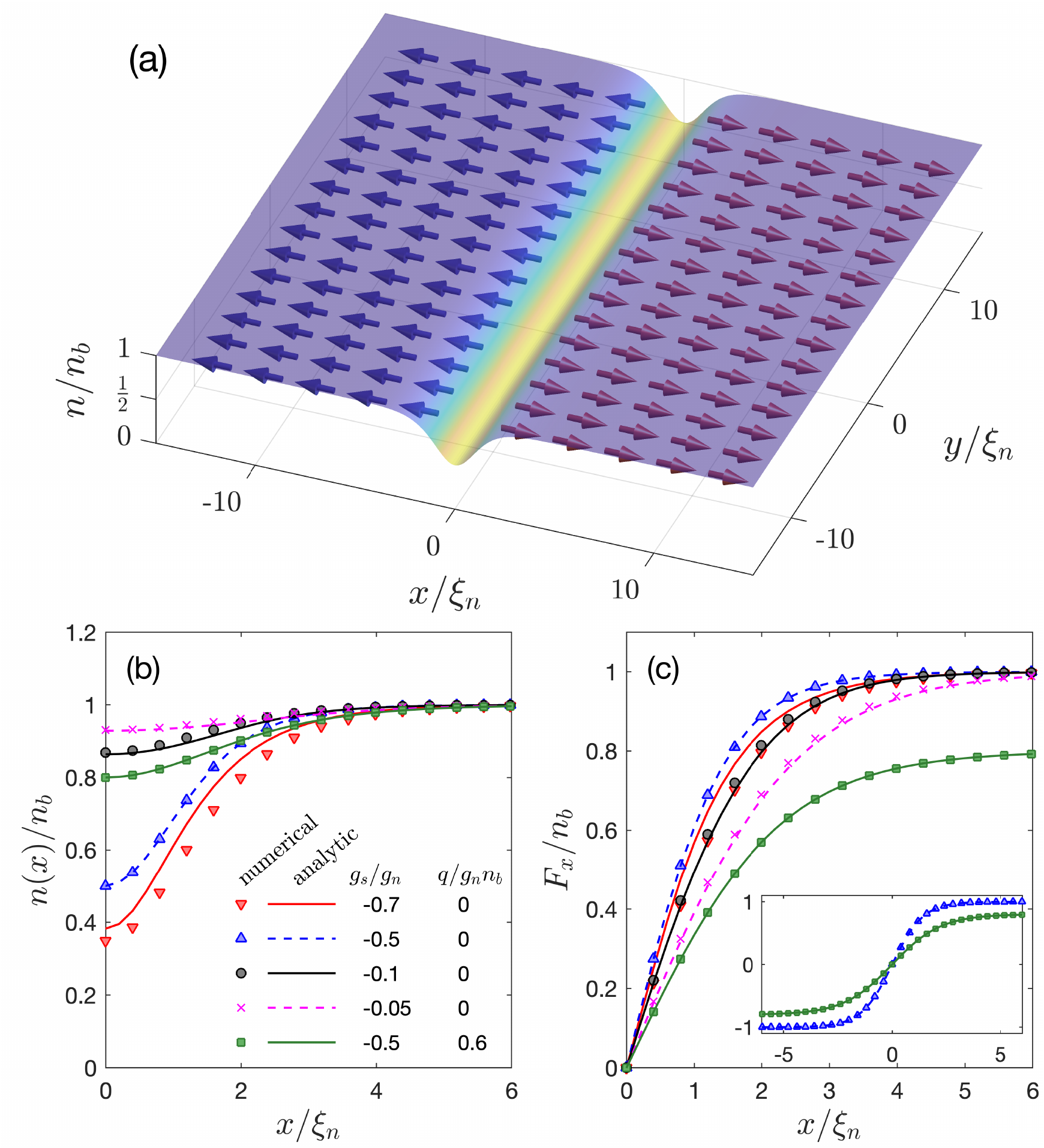}
	\caption{(a) Schematic of a transverse magnetic domain wall (along $y$-axis) in a system of background density $n_b$. The arrows represents the transverse magnetization vector $(F_x,F_y)$, and the background color shows the superfluid density.  A comparison between analytical predictions (lines) and numerical results (symbols) for the (b) density and  (c) $F_x$ spin density for various values of $g_s/g_n$ and $q$.  The inset shows two complete profiles of $F_x$ at the exactly solvable point with $q=0$ and $q\neq 0$, respectively.  Here    $\xi_n=\hbar/\sqrt{M g_n n_b}$ is the density healing length.  	\label{f:domainwall}}
\end{figure}

In this paper we present an analytic solution of a stable  magnetic domain wall in a quasi-2D spin-1 ferromagnetic BEC, characterized by magnetization $\mathbf{F}$ having the typical profile of a dark soliton [Fig.~\ref{f:domainwall}(a)] : a $\pi$ phase  (direction of $\mathbf{F}$)  jump crossing the domain wall and $\mathbf{F}=\mathbf{0}$ at the centre,  breaking the $\mathbb{Z}_2$ symmetry.  In contrast to most domain walls/solitons in BECs, the magnetic domain wall is stable against the snake instability in two dimensions. This is   verified by studying transverse standing waves on these domain walls, finding they oscillate without decay. Instead, the system has a linear dynamic instability driven by modes localized near the domain wall core that cause a growth of the perpendicular components of the magnetization. The resulting spin texture corresponds to a chain of spin vortex anti-vortex pairs along the domain wall. Real time dynamics in the presence of white noise shows that the magnetic domain wall survives.
 Exact solutions are obtained for a large spin-dependent interaction strength $g_s$ with and without magnetic fields. These exact solutions are distinct from two well-known solvable cases: the Manakov regime~\cite{manakov1974theory} ($g_s=0$ in spin-1 BECs) and the magnetic soliton regime (constant number density)~\cite{MDQu2016}. In the absence of magnetic fields, $\textrm{SO}(3)$ spin rotations relate a family of degenerate solutions, and we show that for particular rotations the underlying component wavefunction can map onto a range of solitons and domain walls proposed for binary condensates.
Thus a distinct set of unrelated non-linear excitations are found to be  contained within our solution,  unified by its symmetries.

\section{Formalism for a spin-1 BEC}
The Hamiltonian density of a quasi-2D spin-1 BEC \cite{footnote-1} reads 
 \bea 
\label{Hamioltonian}
	{\cal H}=  \frac{\hbar^2 \left|\nabla \psi\right|^2 }{2M} +\frac{g_n}{2} |\psi^{\dag}\psi|^2+\frac{g_s}{2} |\psi^{\dag} \mathbf{S} \psi|^2 +q \psi^{\dag} S^2_z \psi,
\eea
where the three component wavefunction ${\psi}=(\psi_{+1},\psi_{0},\psi_{-1})^{T}$ describes the condensate amplitude in the  $m=+1,0,-1$ sublevels, respectively. Here $M$ is the atomic mass, $g_n>0$ is the density interaction strength,  $g_s$ is the spin-dependent interaction strength,
$\mathbf{S}=(S_x,S_y,S_z)$ with $S_{\nu=x,y,z}$ being the spin-1 matrices~\cite{footnote-3},  and $q$ denotes the quadratic Zeeman energy. 
The spin-dependent interaction term allows for spin-mixing collisions in which two $m=0$ atoms collide and convert into  $m=+1$ and $-1$ atoms, and the reverse process. 

The  dynamics of the field $\psi$ is given by the Gross-Pitaevskii equations (GPEs) 
$i \hbar \partial \psi/\partial t=\delta {\cal H}/\delta \psi^{\dag} \equiv \cal{L}_{\rm{GP}}\psi$, which in component form is
\begin{subequations}
\bea
\!\!\!\! i\hbar \frac{\partial \psi_{\pm 1}}{\partial t}	&&=\left[H_0+g_s\left(n_0+n_{\pm 1}-n_{\mp 1}\right)+q\right]\psi_{\pm 1}+g_s \psi^2_0 \psi^{*}_{\mp 1}, \\
\!\!\!\! i\hbar \frac{\partial \psi_0}{\partial t}	&&= \left[H_0 +g_s\left(n_{+1}+n_{-1}\right) \right]\psi_0 + 2g_s \psi^{*}_0\psi_{+1}\psi_{-1},  
\eea
\label{GPE2}
\end{subequations}
where $H_0=-\hbar^2\nabla^2/2M +g_n n$, with $n=\sum_mn_m$ and $n_m=|\psi_m|^2$ being the total and component densities, respectively.
Spin-1 BECs  exhibit magnetic order,
e.g., the magnetization $\mathbf{F}\equiv\psi^{\dag} \mathbf{S} \psi$~\cite{magnetization} is the order parameter of ferromagnetic phases $|\mathbf{F}|>0$ for $g_s<0$ ($^{87}$Rb or $^7$Li). In contrast anti-ferromagnetic phases with $g_s>0$ ($^{23}$Na) have $\mathbf{F}=0$. In the absence of magnetic fields, i.e.~$q=0$, ${\cal H}$ is invariant under  $\textrm{SO}(3)$ spin-rotations and the total magnetization $\int d^2\mathbf{r} \ \mathbf{F}$ is conserved. 
 
\section{Dark soliton-like magnetic domain walls} 

For a uniform
ferromagnetic system with total density $n_b$ and at $q=0$, the energy
density ${\cal H}= g_n n^2_b/2+g_s|\mathbf{F}|^2/2$ is minimized for states with
$|\mathbf{F}|=n_b$. The chemical potential is $\mu=(g_n+g_s)n_b$.
We search for a straight line domain wall connecting the two distinct magnetic ground states characterized by $\mathbf{F}=\pm n_b \hat{\mathbf{e}}$, where $\hat{\mathbf{e}}$ is  a 3D unit vector along an arbitrary direction [see Fig.~\ref{f:domainwall}(a)].
For convenience, the domain wall is chosen parallel  to the $y$-axis and the core is located at $x=0$.  We find a solution of the general form\bea
\mathbf{F}\simeq n(x)\tanh\left( {x}/{ \ell}\right)\hat{\mathbf{e}},
\label{MD}
\eea
where $\ell=\hbar/\!\sqrt{4|g_s| M n_b}$.  This result is exact for a particular set of interaction parameters, and a good approximation in general as we discuss further below.
This domain wall is of the Ising type, rather than the Bloch  or N\'eel type,  signified by $\mathbf{F}$ vanishing at the core and changing its sign across the core.  The solution (\ref{MD}) has the characteristic profile of dark soliton and we refer to it as dark soliton-like magnetic domain wall (MDW).  This domain wall is in magnetic order but not in the superfluid order, i.e.~the superfluid density $n(x)$ does not vanish, but has a dip at the core to minimize the energy.

\subsection{Exact solutions}
When the width of the density dip coincides with $\ell$, occurring at $g_s=-g_n/2$~\cite{footnotescattering},  Eq.~\eqref{GPE2} admits an exact solution
\bea
\mathbf{F}(x)=n_b \tanh\left(\frac{x}{2 \ell}\right)\hat{\mathbf{e}}, \quad n(x)=n_b \left[1-\frac{1}{2}\sech^2\left(\frac{x}{2\ell}\right)\right].
\label{MDexact}
\eea
 
This system has a  $\textrm{SO}(3)$ symmetry which relates a continuous family of degenerate  MDW solutions connected  by $\textrm{U}(1)$  gauge and spin ${\cal U}(\alpha,\beta,\tau)=e^{-i \alpha S_{z}}e^{-i \beta S_y} e^{-i \tau S_z}$ rotations, where $\{\alpha,\beta,\tau\}$ are the Euler angles.
We illustrate three members of this family 
in Table~\ref{table}: (i)
For the case of an $F_x$ domain wall [i.e.~$\hat{\mathbf{e}}=\hat{\mathbf{x}}$], the underlying wavefunctions can have two distinct vector soliton profiles, and the corresponding stationary GPE can be mapped onto that of a miscible binary BEC. (ii) A Sine-Gordon type soliton (SGS) of the phase difference $\theta_{d}\equiv\theta_{\pm 1}-\theta_{0}$, where  $\psi_m=|\psi_m|e^{i\theta_m}$.
A SGS has been predicted to exist in a coherently-coupled binary BEC, with dynamics mimicking  processes in quantum chromodynamics~\cite{Son2002}.  
Here the SGS can be produced by a spin rotation of the vector soliton in Table~\ref{table} and the nonlinear spin-mixing interaction provides the necessary couplings between the component phases, having the advantage that no external fields are required~\cite{footnoteSGS}. (iii) For an $F_z$ domain wall, the corresponding wavefunction coincides with a (density) domain wall of an immiscible binary BEC~\cite{Ao1998a}.

\begin{table*}[!t]
	\centering 
	\begin{tabularx}{\textwidth}{c|c|c|c}	
		\hline\hline
		($\alpha$, $\beta$, $\tau$) &type-I: 0; \quad type-II: ($\pi/2$, $-\pi/2$, $-\pi/2$) & ($-\pi/2$, $-\pi/4$, $-\pi/2$) & ($\pi/2$, $\pi/2$, $0$) \\
		\hline
	$\textrm{U(1)}$ &type-I: 1; \quad type-II: $e^{i \pi/2 }$ & $e^{-i \pi/4 }$ & $e^{i 3\pi/2}$ \\
		\hline
	\multicolumn{1}{p{0.01\textwidth}|}{\footnotesize $\psi \big|_{g_s=-g_n/2}$} &I: $\psi_{\pm 1}=\sqrt{n_b}/2 \tanh\left(x/2\ell\right)$; $\psi_0=\sqrt{n_b/2}$ &$\psi_{\pm1}=\sqrt{n_{\pm1}}e^{i\theta_d/2},\psi_{0}=\sqrt{n_0}e^{-i\theta_d/2}$ &$\psi_{\pm 1}=\sqrt{n_b}/2\left[1\mp\tanh\left(x/2\ell\right)\right]$     \\ 
		&II: $\psi_{\pm 1}=\sqrt{n_b}/2,\psi_{0}=\sqrt{n_b/2} \tanh(x/2\ell)$ & $\theta_d(x)=2\arctan e^{x/\ell}$, $2n_{\pm1}=n_0=n/2$ &$\psi_{0}=0$\\
		\hline
		$\mathbf{F}$ & $F_x=n_b \tanh\left(x/2 \ell\right)$ & $F_x=n_b \tanh\left(x/2 \ell\right)$ & $F_z=-n_b \tanh\left(x/2 \ell\right)$  \\
		\hline 
		GPE & $0=\left[H' + 2g_nn_{\pm 1} +(g_n+2g_s) n_0\right]\psi_{\pm 1}$ &$0=\frac{\hbar^2}{2M}\partial_x (n \partial_x \theta_d)  +g_s n^2\sin (2\theta_d)$ & $0=\left[H'+(g_n+g_s) n_{+1}+(g_n-g_s)n_{-1}\right]\psi_{+1}$  \\
		&$0= \left[H' +g_nn_0 +2(g_n+2g_s)n_{\pm 1} \right]\psi_{0\phantom{+}}$ & $0 	=\frac{\hbar^2}{2M}\left[\frac{1}{2}(\partial_x\theta_d)^2-\frac{2}{\sqrt{n}} \partial^2_x \sqrt{n}\right] $ &  $0=\left[H'+(g_n+g_s) n_{-1}+(g_n-g_s)n_{+1}\right]\psi_{-1}$  \\
		&$H'=-\frac{\hbar^2}{2M}\partial_x^2-\mu$ &$\phantom{0=}+2[n(g_n+g_s \cos^2\theta_d)-\mu]$ &\\[1pt]
		\hline
			\multicolumn{1}{p{0.05\textwidth}|}{\footnotesize Related systems} & \multicolumn{1}{p{0.23\textwidth}|}{\footnotesize Vector soliton of a three-component BEC and a miscible binary BEC} &  \multicolumn{1}{p{0.25\textwidth}|}{\footnotesize Sine-Gordon type soliton, also realized in a   coherently-coupled binary BEC }
		& \multicolumn{1}{p{0.29\textwidth}}{\footnotesize  Density domain wall of an immiscible binary BEC} \\
		\hline\hline
	\end{tabularx}
	\caption{Component representation of the MDW after various spin rotations. Vector soliton sector: type-I vector soliton is chosen as a reference point. In this presentation, the reduced GPEs are related to a miscible binary system and becomes decoupled at $g_s=-g_n/2$, allowing the exact solution. SGS: $\theta_d$ satisfies the Sine-Gordon equation. Binary domain wall sector: the reduced GPEs describe an immiscible binary system and the corresponding exact solution coincides with a solution discussed in a different context~\cite{Malomed1990}. 
	}
	\label{table}
\end{table*}

In the context of binary BECs, the vector solitons, the SGS and the density domain wall are unrelated. In a spin-1 BEC, these distinct nonlinear excitations are unified by spin rotations of our MDW solution. 
With inadequate degrees of freedom and symmetries, such connection can not be made within the binary BEC~\cite{footnoteManakov, StamperRMP}. However it is important to note that the dynamics and stability properties of the MDW reveal the spin-1 nature and exhibit distinct behaviors from related excitations in binary systems (see below). A recent study on magnetic solitons in anti-ferromagnetic BECs has also explored the role of the rotational symmetry~\cite{chai2020magnetic}. 
  
\subsection{Away from the exactly solvable point}
Away from the exactly solvable point we develop a self-consistent asymptotic analysis of the stationary GPEs at $x\gg \ell$, combined with an account of the local core structure, and we find an accurate  approximate form for the density
\begin{subnumcases}{\label{SSDW} \hspace*{-0.5cm}\frac{n(x)}{n_b} \simeq}
	\frac{ \cosh(x/\lambda \ell)}{a_1 \cosh(x/\lambda \ell)+a_1^2b_1}+1-\frac{1}{a_1},  &  \hspace*{-0.125cm} $g_s<-\frac{g_n}{5}$ \label{SSDW1}\\
	1+\frac{4b_1 \, g_s}{2(g_n+g_s)\cosh^2(x/\ell)+g_1},  & \hspace*{-0.125cm} $-\frac{g_n}{5}<g_s$\label{SSDW2}
\end{subnumcases} 
where $\lambda=\sqrt{-g_s/(g_n+g_s)}$, 
$a_1=- (2 g^2_n +2 g_n g_s-g_s^2)/(3g_s(g_n + g_s))$, $b_1=3 (g_n + g_s)/(2 g_n+g_s)$,
and 
$g_1=2 b_1(2 g_n - 5 g_s)/3$.

In the following we show the procedure to obtain Eqs.~\eqref{SSDW}.  Specializing to the SGS  (see Table \ref{table}) we work with the hydrodynamical variables $\{n,n_d,\theta_d,\theta_s\}$, where $n_d=2n_{\pm 1}-n_0$, $\theta_s=\theta_{\pm 1}+\theta_{0}$, $\theta_{+1}=\theta_{-1}$ and $n_{+1}=n_{-1}$. 
For a stationary state, the total number current 
$\mathbf{J}_{n}=\hbar^2/(2M)\left(n \bm{\nabla} \theta_s + n_d \bm{\nabla} \theta_d\right)$
should vanish. Apparently $\theta_s=0$ and $n_d=0$ solve $\mathbf{J}_{n}=0$, and for this case Eqs.~\eqref{GPE2} [or Eqs.~\eqref{hGPE}] reduce to 
\begin{subequations}
	\label{hGPE2}
	\bea
	\label{SS1}
	0&&=\frac{\hbar^2}{2M}\partial_x (n \partial_x \theta_d)  +g_s n^2\sin (2\theta_d), \\
	\mu &&=\frac{\hbar^2}{4M}\left[\frac{(\partial_x\theta_d)^2}{2}-
	\frac{2 \partial^2_x \sqrt{n}}{\sqrt{n}}\right]+ {g_s n}\cos^2 \theta_d  +g_n n. 
	\label{SS2}
	\eea
\end{subequations} 
As shown in Table \ref{table}, at $g_s=-g_n/2$, 
\bea
\theta_d(x)=2\arctan e^{x/\ell},
\label{exactsolution}
\eea
where $\ell=\hbar/\sqrt{4 |g_s|Mn_b }$ as introduced earlier.  Away from the exactly solvable point we assume that the expression of $\theta_d(x)$ in Eq.~\eqref{exactsolution} remains a good approximation.
In other words, $\mathbf{F}/n(x)$ is assumed to take the same form as at the exactly solvable point. The reason for this will become clear later.

Let us examine the asymptotic form of Eq.~\eqref{SS2} far away from the core $x=0$. Assuming that
$g(x)\equiv [n(x)-n_b]/(4 n_b)$
decays  slower than $(\partial_x \theta_d)^2 \sim e^{-2 x/\ell}$ for large $x\gg \ell$ (there is no solution for $g(x)$ decaying faster than $e^{-2 x/\ell}$), in the large $x$ limit, the dominant part of Eq.~\eqref{SS2} reads
\bea
\left(g_n+g_s\right)g(x)+\ell^2  g_s g''(x)=0,
\eea
having a solution $g(x\gg \ell)\sim e^{-x/\ell_d}$, where $\ell_d=\lambda \ell$ is the effective density length scale. Combining the asymptotic behavior of $n(x)$ at large $x$ and $n(x)$ being an even function, it is natural to propose the ansatz~\eqref{SSDW1}. The coefficients  $a_1$ and $b_1$ are introduced to adjust the core structure and are determined by requiring that $n(x)$ satisfies  Eqs.~\eqref{hGPE2} to leading order as $x\rightarrow 0$. The working assumption to obtain Eq.~\eqref{SSDW1} is $\ell_d>\ell (\lambda>1/2)$, implying $g_n+5g_s<0$ which sets the parameter range for the solution in Eq.~\eqref{SSDW1} to be applicable.  
This regime includes the exactly solvable point, $g_s=-g_n/2$ with $\lambda=1$, where $\ell_d= \ell$, and a single length scale describes the spin and density character of the MDW [here (\ref{SSDW1}) reduces to Eq.~(\ref{MDexact})]. In this strong spin interaction regime (\ref{SSDW1}), the density variation near the core is important.  The excitation breaks down at $g_s+g_n=0$ which is the parameter boundary of the ferromagnetic phase\cite{StamperRMP,KAWAGUCHI12}.

In the opposite limit, where $|g_s/g_n|\ll 1$,  the quantum pressure term $\sim \partial^2_x \sqrt{n}/\sqrt{n}$ becomes less important and can be neglected. Hence Eq.~\eqref{SS2} becomes an algebraic equation of $n(x)$ with solution given Eq.~\eqref{SSDW2}. The parameters $b_1$ and $g_1$ are introduced to solve Eqs.~\eqref{hGPE2} near the core $x=0$ to leading order. The crossover to the weak spin interaction regime (\ref{SSDW2})  occurs at $g_n+5g_s=0$ where $\lambda=1/2$, given by matching the density widths $\ell/2$ and $\ell_d$.  For comparison we calculate numerical MDW results using a gradient flow method \cite{Lim2008a,Bao2008a}.  The analytic and numerical results in Figs.~\ref{f:domainwall}(b) and (c) show excellent agreement.

Let us now provide a self-consistent reasoning to  explain why $\theta_d(x)$ in Eq.~\eqref{exactsolution} serves a good approximation in the whole parameter range. First of all,  it captures the main feature of the domain wall in the strongly interacting regime where the exact solution Eq.~\eqref{exactsolution} is found.  On the other hand,  in the weak interaction limit ($|g_s/g_n| \ll 1$) the density $n$ can be approximated as a constant ($n\simeq n_b$) and the energy density becomes
\bea
{\cal H}
=-\frac{\hbar^2n}{8M}|\nabla \theta_d|^2+\frac{g_n }{2} n_b - \frac{1}{2} g_sn_b^2\cos^2 \theta_d. 
\eea
A local minimum of the energy density, determined by $\delta {\cal H}/ \delta \theta_d=0$, leads to the elliptic sine-Gordon equation 
$\hbar^2/(2M) \nabla^2(2\theta_d) + 2n_bg_s \sin (2\theta_d)=0$, having the solution $\theta_d= 2 \arctan e^{x/\ell}$.

Since the magnetization vanishes at the core, there is no spin current across the MDW.  However, the nematic tensor current is nontrivial~\cite{footnotecurrent}. The component number currents vary for different degenerate states. For example, with reference to the states in Table \ref{table}: the component currents are zero for the vector soliton, while for SGS  there are internal currents near the core that behave analogously to Josephson currents~\cite{barone1982physics}. 

\section{Finite magnetic fields}
A magnetic field along the $z$-axis breaks the $\textrm{SO}(3)$ symmetry and the degeneracy of states presented in Table~\ref{table} is lifted. For $q>0$ the ground state magnetization prefers to be transverse, realizing an easy-plane ferromagnetic phase that possesses a remnant $\textrm{SO}(2)$ symmetry  \cite{StamperRMP,KAWAGUCHI12}. Here the SGS and the binary density domain wall are no longer stationary solutions.   
The type-I vector soliton is energetically favored, and exists, with some modifications, in the whole easy-plane phase ($0<q<2|g_s|n_b$). 
At $g_s=-g_n/2$, the exact solution is
\bea
\mathbf{F}&=&n_b\sqrt{1-\tilde{q}^2} \tanh\left(\frac{x}{2\ell_q}\right)\hat{\mathbf{e}}_{\rho},\\ n(x)&=&n_b\left[1-\frac{1-\tilde{q}}{2}\sech^2\left(\frac{x}{2\ell_q}\right)\right],
\eea
where $\tilde{q}=-q/2g_sn_b$, $\ell_q= \!\ell/\sqrt{1-\tilde{q}}$ and $\hat{\mathbf{e}}_{\rho}$ is a unit vector in the $xy$-plane.  The corresponding wavefunction reads
$\psi_{\pm 1}=\sqrt{n_b(1-\tilde{q})/4}\tanh\left(x/2\ell_d\right)$,
and $\psi_{0}=\sqrt{n_b (1+\tilde{q})/2}$. An example of a $q\ne0$ result ($\hat{\mathbf{e}}_{\rho}=\hat{\mathbf{x}}$) is shown in Fig.~\ref{f:domainwall}.

\section{Standing waves}
\begin{figure}[htp] 
	\centering
		\includegraphics[width=0.47\textwidth]{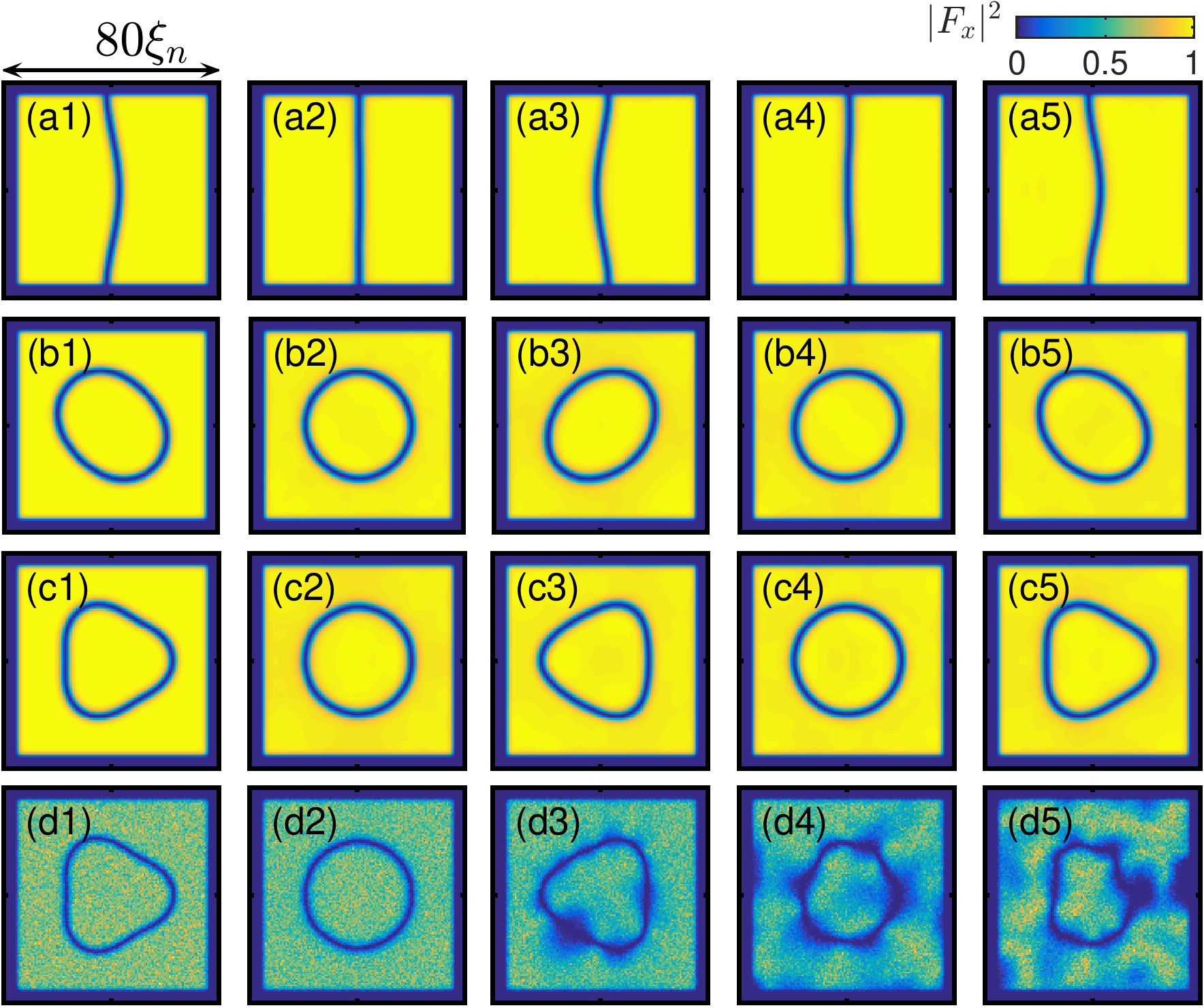}
	\caption{One period of  evolution for standing wave deformations of a MDW confined by hard wall potentials in a square domain $x,y\in[-L,L]$ with $L=40 \xi_n$.  The equilibrium configurations are shown in the second and forth columns. (a1)-(a5) A standing wave on an open MDW with free end-points attached on the hard wall boundaries; the initial configuration (MDW core location) is $x= A \cos(\pi y/L)$, $y\in [-L,L]$ with $L=40 \xi_n$ and $A=2.5 \xi_n$. (b1)-(b5), (c1)-(c5) Standing waves on a closed MDW. The initial configurations are determined by $R_0-\sqrt{x^2+y^2}+A \sin[ s \arctan(y/x)]=0$ where $R_0=L/2$, $s=2$ (dipole mode) for (b1) and $s=3$ (triple mode) for (c1). Oscillation periods  $T$ are: (a) $T \simeq 620t_0$; (b) $T \simeq 620t_0 $; (c) $T \simeq 300t_0$, where $t_0=\hbar/g_nn_b$ and  $g_s/g_n=-0.1$.  (d1)-(d5) Results with white noise added to the initial condition of (c1), causing a $\!\sim\!1$\% increase in particle number.  Such stability also stands for the open domain walls. Note that for (a) (b) and (c) $|F_x|=|\mathbf{F}|$.
	 \label{f:realtimedynamics}}
\end{figure}

\begin{figure}[htp]
	\centering
	\includegraphics[width=0.47\textwidth]{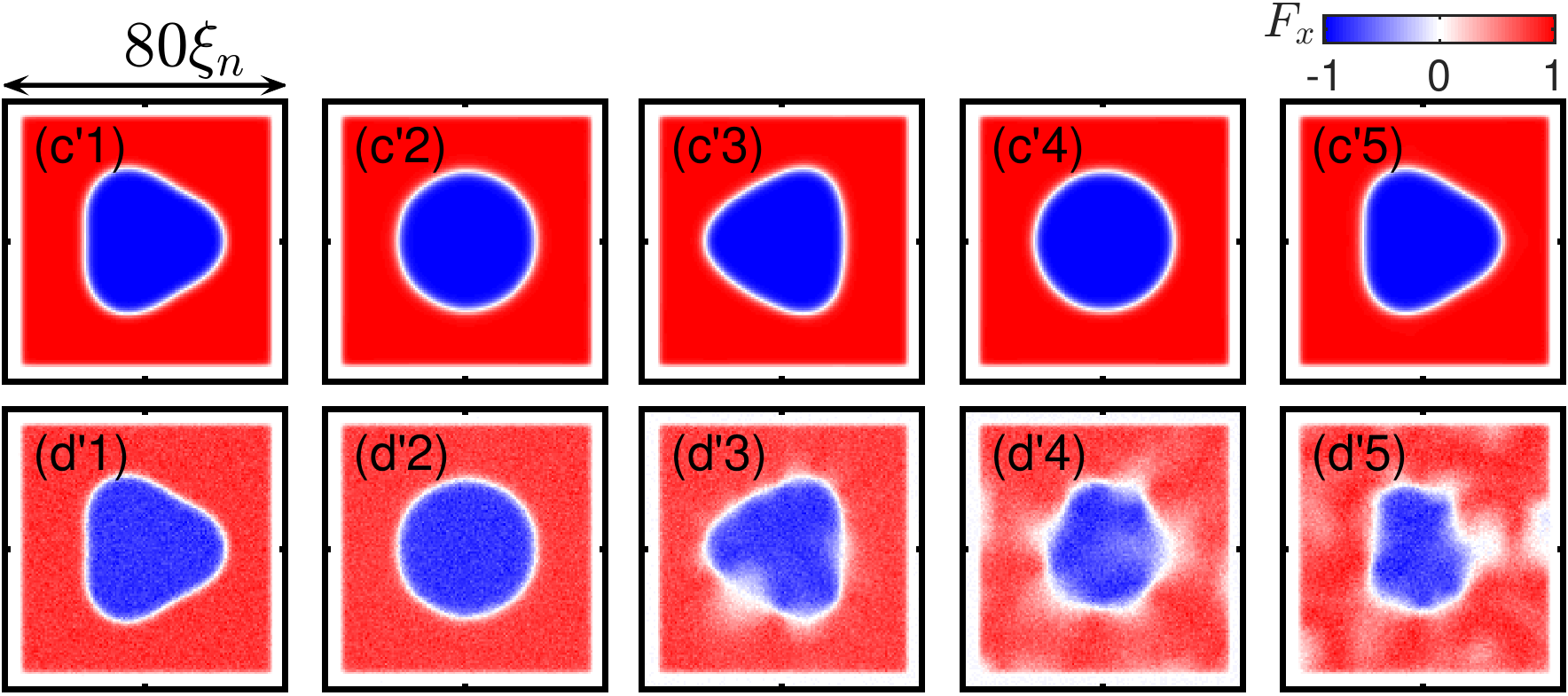}
	\caption{
		\label{f:realtimedynamicsphase} The transverse magnetization $F_x$ during the time evolution. $(\text{c}'1)$-$(\text{c}'5)$ and $(\text{d}'1)$-$(\text{d}'5)$ correspond to Fig.~\ref{f:realtimedynamics}(c1)-(c5) and (d1)-(d5), respectively.} 
\end{figure}

A conspicuous feature of the 2D dynamics of the MDW is that it is  stable against transverse deformations, strikingly different from other domain walls/solitons~\cite{kuznetsov1988instability,Shlyapnikov1999,DSdecayExp,Son2002,Kasamatsu2019,Gallemi2019,MDQu2016,MSexp1,MSexp2,shin2019} which decay unavoidably via snake instability.  We  consider easy-plane domains with $\mathbf{F}$ along the $x$-axis and two fundamental static MDW geometries in the $x$-$y$ plane for $q=0$: closed circle and open straight line with endpoints attached on the boundaries~\cite{footnotefreeends}  (see Fig.~\ref{f:realtimedynamics}).
We excite standing waves on these static MDWs  by deforming them transversely. The subsequent time evolution shown in Figs.~\ref{f:realtimedynamics}(a)-(c) is periodic and resembles harmonic modes vibrations~\cite{footnotelargedeformation}. Our simulations~\cite{Symes2016a} also show that the standing waves persist without decay,  combined with  internal spin-exchange dynamics between components of the wavefunction (see Fig.~\ref{f:spinexchange} in Appendix \ref{appendixb}).  During the time evolution $F_y$ and $F_z$ remain zero, the magnetization conservation manifests itself as a geometrical constraint of the domain wall motion: the area enclosed by the domain wall remains unchanged.  There is no spin current crossing  the MDW. The enclosed regions form  magnetic bubbles, inside which the magnetization $F_x$ has the opposite orientation from the outer one and such feature remains in the presence of noise (see Fig.~\ref{f:realtimedynamics} and Fig.~\ref{f:realtimedynamicsphase}). Consequently, propagating open MDWs and expanding/shrinking closed MDWs are prohibited, becoming possible when applying magnetic fields along the $z$-axis.

\section{Dynamical instability} 
Here we systematically study the stability of the MDW by means of Bogoliubov-de Gennes equations (BdGs).  
Let us consider a straight infinitely long MDW along the $y$-axis located in the middle of a slab of width $L_s\gg \ell$.
Denoting the stationary MDW as $\psi_s$,  we consider a perturbation   $\delta \psi=u(\mathbf{r})e^{-i\omega t}+v^{*}(\mathbf{r})e^{i \omega^{*} t}$. Linearizing about $\psi=\psi_s+\delta \psi$ in Eq.~\eqref{GPE2} yields the  BdG equations  
\bea
\hbar \omega  \left( {\begin{array}{cc}
		u\\
		v\\
\end{array} } \right)=
\left( {\begin{array}{cc}
		{\cal L}_{\rm GP}+X-\mu & \Delta \\
		-\Delta^{*} & -({\cal L}_{\rm GP}+X-\mu)^{*} \\
\end{array} } \right)
\left( {\begin{array}{cc}
		u\\
		v\\
\end{array} } \right),
\label{BdG}
\eea
where the stationary wavefunction satisfies  ${\cal L}_{\rm GP}\psi_s=\mu \psi_s$, $X= g_s\sum_\nu S_\nu\psi_s\psi_s^{\dagger}S_\nu+ g_n \psi_{s}\psi^{\dagger}_s $, $\Delta=g_n \psi_s \psi^{T}_s+g_s \sum_{\nu} (S_\nu \psi_s)(S_\nu\psi_s)^{T}$ and $\mu=(g_n+g_s)n_b+q/2$.  
\begin{figure}[htp] 
	\centering
	\includegraphics[width=0.43\textwidth]{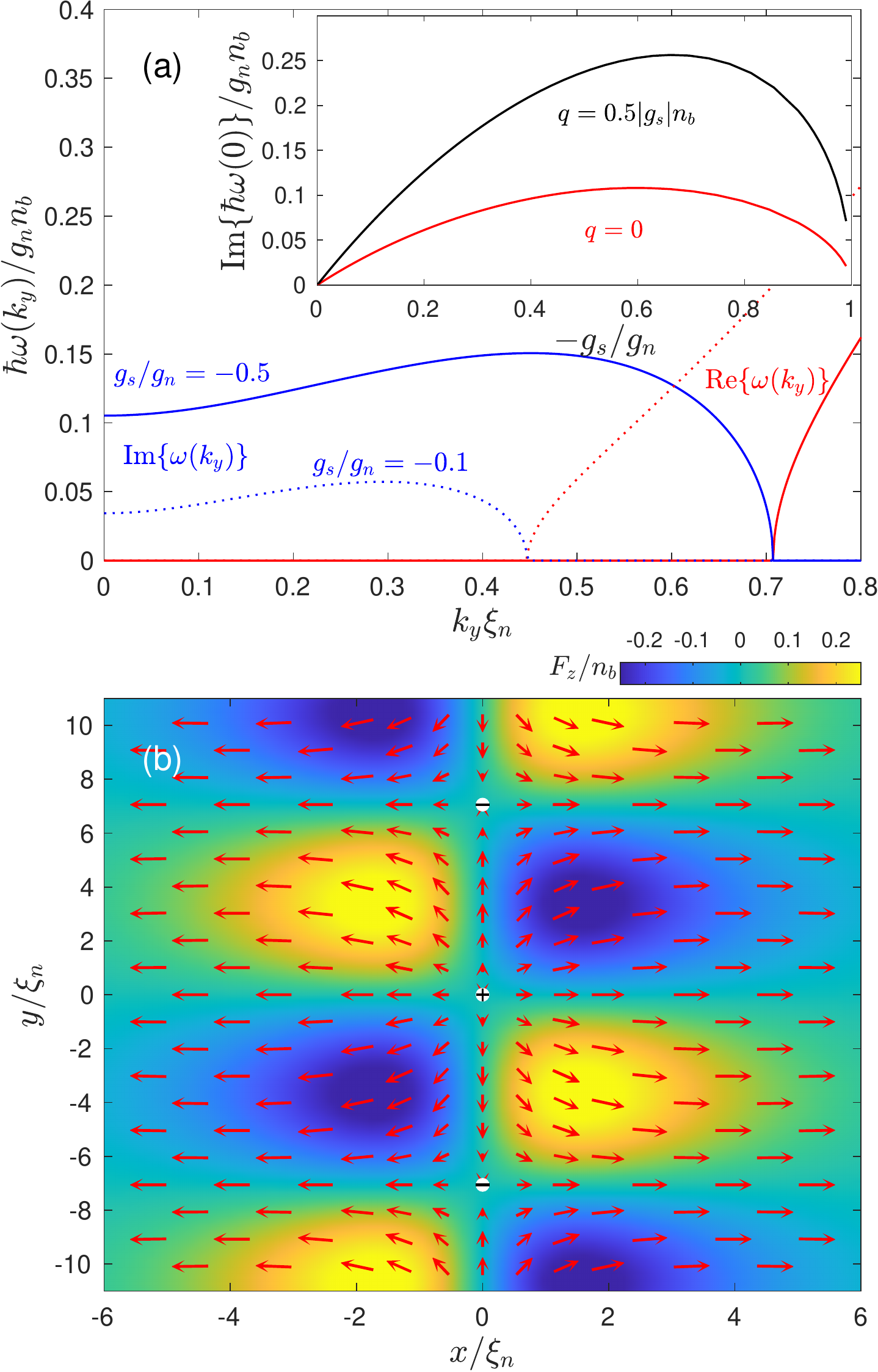}
	\caption{Unstable spin-twist modes. (a) Spectrum of the unstable modes for $q=0$ and two values of $g_s/g_n$. For $g_s/g_n=-1/2$ the bifurcation point ($\omega=0$) occurs exactly at $k_y\xi_n=1/\sqrt{2}$. Insets shows the magnitude of the long wavelength instability as $g_s/g_n$  varies. (b) The spin-texture created by the unstable mode~\cite{footnoteBdG} at $k_y\xi_n\simeq0.445$ where the maximum imaginary frequency is reached for $q=0$ and $g_s/g_n=-1/2$. White circles with $+$ and $-$ indicating positive and negative circulation spin-vortices, respectively. The red arrows and the background color represent transverse magnetization field $(F_x,F_y)$ and longitudinal magnetization  $F_z$, respectively.
		\label{f:imaginary1}  } 
\end{figure}     
The translational symmetry along $y$ allows us to parameterize the perturbations with the wave-vector $k_y$ as $u(\mathbf{r})=u(x) e^{ik_y y}$ and $v(\mathbf{r})=v(x) e^{ik_y y}$.  We numerically solve Eq.~\eqref{BdG} with Neumann boundary conditions at $x$-axis boundaries~\cite{footnote2}, and find two modes with an imaginary energy [Fig.~\ref{f:imaginary1}(a)], marking a  dynamical instability in the system (a mode that grows exponentially with time).

Different from the snake instability~\cite{Shlyapnikov1999},  the imaginary part of the excitation energy  $\rm Im(\omega)$ does not vanish as  $k_y\to0$, but instead  approaches a finite value  [Fig.~\ref{f:imaginary1}(a)],  implying that this instability also exists in 1D. 
$F_x$ is unchanged as the unstable mode grows, however it causes the unmagnetized core of the MDW to develop a $F_y$-texture of wavelength $\pi/k_y$. This corresponds to the formation of a chain of ``magnetic vortex'' cores~\cite{footnote3} at the  nodes of this texture [Fig.~\ref{f:imaginary1}(b)].

The $k_y$ range of unstable modes and the magnitude of the imaginary energy is largest at intermediate values of $g_s/g_n$, and increases with increasing $q$  [see inset to Fig.~\ref{f:imaginary1}(a)].
Based on the magnetic texture created by the unstable mode, we refer to it as spin-twist instability. In dynamics the growth of this instability leads to spin waves of $F_y$ and $F_z$ while the topological structure of the MDW in $F_x$ remains unchanged, consistent with the noisy dynamics observed in Fig.~\ref{f:realtimedynamics}(d).  Note that this characteristic feature does not rely on the conservation law of magnetization, and holds in the presence of magnetic fields ($q>0$).  The growth of $F_y$ and $F_z$ could be stabilizing the domain wall by absorbing the perturbations, in analogy to a buffering effect.

\section{Conclusion and outlook}
 We found a novel magnetic domain wall in a quasi-2D ferromagnetic spin-1 BEC that 
is stable against the snake instability and  white noise.   Along with the exact solutions, an accurate analytic solution applicable to the whole ferromagnetic phase has also been developed. 
Through the underlying symmetries of the spin-1 system, we have shown that various distinct nonlinear structures such as the Sine-Gordon soliton, vector solitons and  an immiscible binary density domain wall occurring in unrelated binary systems are unified into the magnetic domain wall. 
Our findings open a new possibility to study rich 2D dynamics of domain wall solitons, could be important for determining the universality class of the ferromagnetic phase transition of 2D spin-1 BECs at finite temperature~\cite{James2011,KobayashiBKT},  coarsening dynamics involving both spin order and superfluid order~\cite{Andreane2017a,Prufer2018a} and dynamics of stretched polar-core vortices~\cite{polarcorevortexLewis,polarcorevortexTurner}.

It will be feasible to observe magnetic domain walls in current experiments with ferromagnetic spinor BECs.
The necessary techniques  for manipulating the spin degrees of freedom of a spin-1 BEC \cite{MSexp2}, and for non-destructively measuring its spin dynamics \cite{Higbie2005} have already been demonstrated.  Coupled with a planar or flat-bottom optical trap (e.g.~\cite{Dalibard2015,Gauthier16}) opens the possibility for investigating of domain wall dynamics. Most work with ferromagnetic spin-1 BECs to date has been conducted with $^{87}$Rb which has $-g_s/g_n\sim10^{-2}$ and is in the weakly spin-interacting regime. However, recently a   $^7$Li spin-1 BEC has been prepared with $-g_s/g_n\sim0.5$ \cite{Huh2020a}, thus in the strong spin interacting regime close to the exactly solvable point.

\section{Acknowledgment}
We thanks M. Antonio, T. Luke, L. A. Williamson, Russell Bisset and Danny Baillie for useful discussions. 
X.Y. acknowledges the support from NSAF through grant number U1930403. P.B.B acknowledges support from the Marsden Fund of the Royal Society of New Zealand.

\appendix

\section{Spin-1 Gross-Pitaevskii equations in hydrodynamical variables}

In terms of  the hydrodynamical variables $\{n,n_d,\theta_d,\theta_s\}$, the stationary GPE for $q=0$ becomes   
\begin{widetext}
	\begin{subequations}
		\label{hGPE}
		\bea
		\label{fEQ1}
		0 &&=-\frac{\hbar^2}{2M}\nabla\cdot (  n \nabla \theta_s + n_d \nabla \theta_d), \\
		\label{fEQ2}
		0	&&=\frac{\hbar^2}{2M}\nabla \cdot(n \nabla \theta_d+ n_d \nabla \theta_s)  +g_s (n^2-n^2_d)\sin (2\theta_d), \\
		\label{fEQ3} 
		0 	&&=-\frac{\hbar^2}{2M}\left[\frac{1}{\sqrt{n+n_d}}\nabla^2 \sqrt{n+n_d}+\frac{1}{\sqrt{n-n_d}}\nabla^2 \sqrt{n-n_d}-\frac{1}{2}(|\nabla \theta_s|^2+|\nabla \theta_d|^2)\right]+ g_s n  \cos (2\theta_d) + (g_s+2 g_n) n  - 2\mu , \\
		\label{fEQ4}
		0	&&=\frac{\hbar^2}{2M}\left[\frac{1}{\sqrt{n+n_d}}\nabla^2 \sqrt{n+n_d}-\frac{1}{\sqrt{n-n_d}}\nabla^2 \sqrt{n-n_d}-\nabla \alpha \cdot \nabla \beta\right]+ g_s n_d  \cos (2\beta) +g_s n_d, 
		\eea
	\end{subequations}
\end{widetext}
where $n_d=2n_{\pm 1}-n_0$, $\theta_s=\theta_{\pm 1}+\theta_{0}$, $\theta_{d}=\theta_{\pm 1}-\theta_{0}$, $n=n_{+}+n_{-}+n_{0}$, $\mu=(g_n+g_s)n_b$ is the chemical potential and $n_b$ is the  ground state total number density. 
Here we assume that $\theta_{+1}=\theta_{-1}$ and $n_{+1}=n_{-1}$.

\section{Real time evolution}
\label{appendixb}
\begin{figure}[H] 
	\centering
	\includegraphics[width=0.43\textwidth]{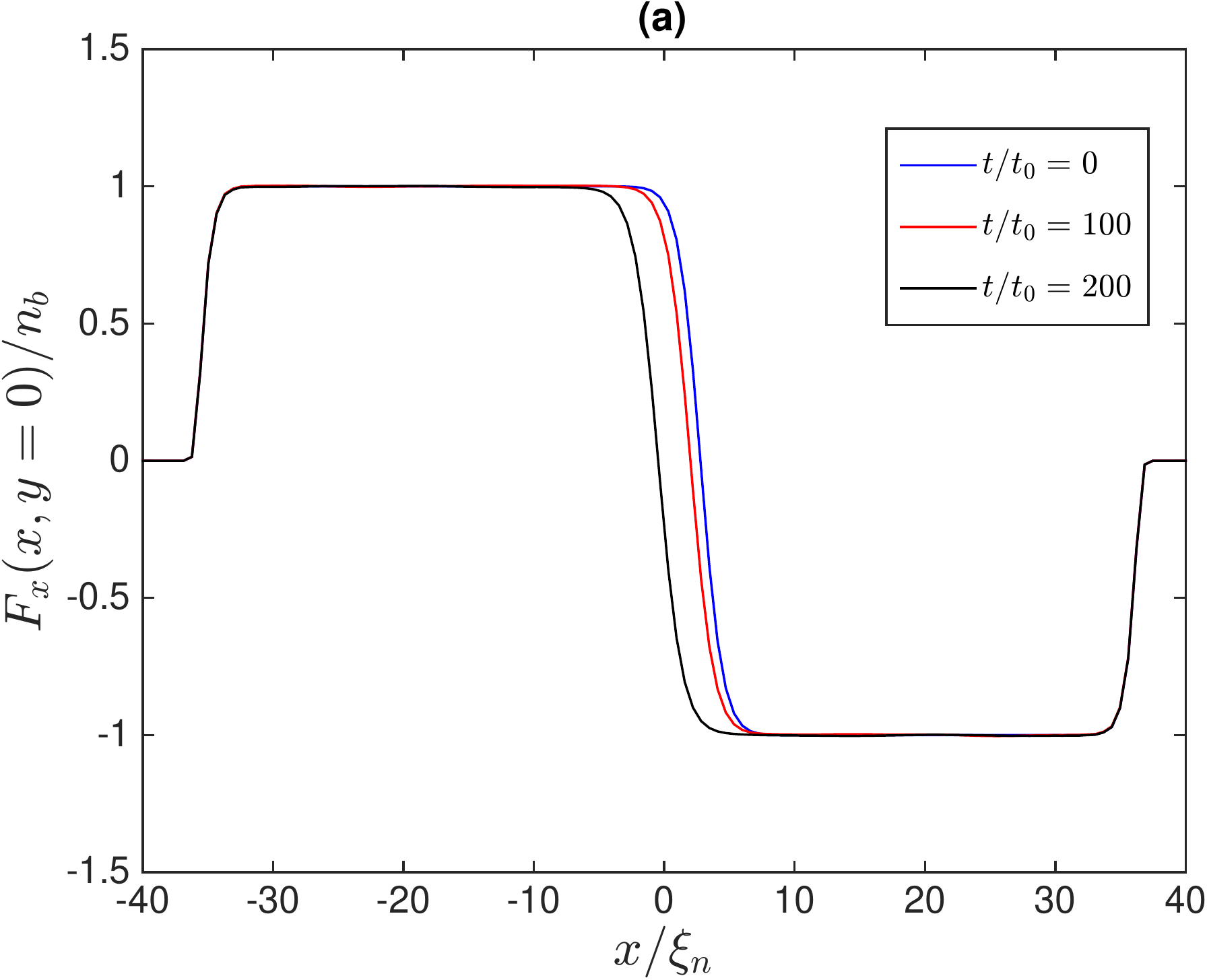}
	\includegraphics[width=0.43\textwidth]{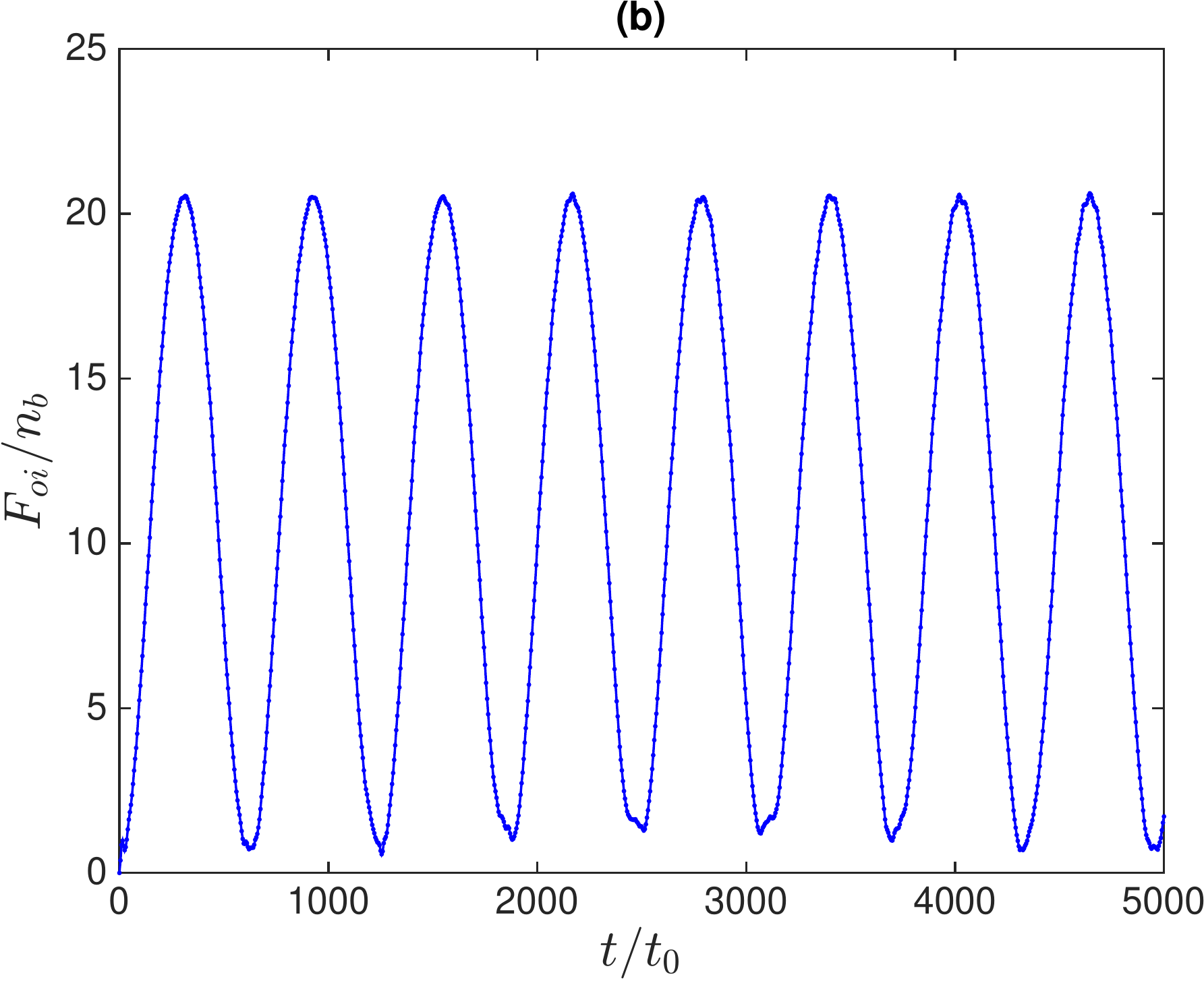}
	\caption{
		\label{f:section} (a) shows the spin-density cross section $F_{x}(x,y=0)$ of the open MDW shown in Fig.~\ref{f:realtimedynamics}(a) at different times.  The soliton-like profile of the magnetization is preserved during the motion. (b) shows the periodic behavior of the overlap function for the open domain wall configuration [Fig.~\ref{f:realtimedynamics}(a)], demonstrating that the standing wave on the MDW persists without decay over long time periods. } 
\end{figure}

Here we present further evidence of the stability. Figure~\ref{f:realtimedynamicsphase} shows that the topological nature of the MDW, i.e.~the $\pi$ phase jump across the core, is well preserved during the domain wall motion. This can be also seen in Fig.~\ref{f:section}(a) which shows the profile of the transverse magnetization $F_{x}(x,y=0)$ at different times.

In our simulations, the box potential takes the following form
\bea
\hspace{-0.6cm}V(x,y)=V_0\left\{2-\tanh\left[\left(\frac{L}{c}-|x|\right)b\right]-\tanh\left[\left(\frac{L}{c}-|y|\right)b\right]\right\},
\eea
where $x,y\in [-L,L]$, $L$ is the box size, $V_0 \gg \mu$ is the hight of the potential barrier, $b \sim \xi^{-1}_n$ is the width and $c$ should be chosen slightly greater than one. 

The standing wave excitation on the MDW can last a long time without decay. In order to quantify this property, we introduce an overlap function  
\bea
O(t)\equiv \left(\int d^2 \mathbf{r} \left|F_{x}(\mathbf{r},t)-F_{x}(\mathbf{r},0)\right|^2\right)^{1/2}
\eea
that measures the overlap of transverse magnetization profile at time $t$ with its initial profile.  
Figure \ref{f:section}(b) shows the periodic behavior of  $O(t)$ for the open domain wall [Fig.\ref{f:realtimedynamics}(a)], revealing that the standing wave on the MDW persists without decay.

\begin{figure}[htp] 
	\centering
	\includegraphics[width=0.47\textwidth]{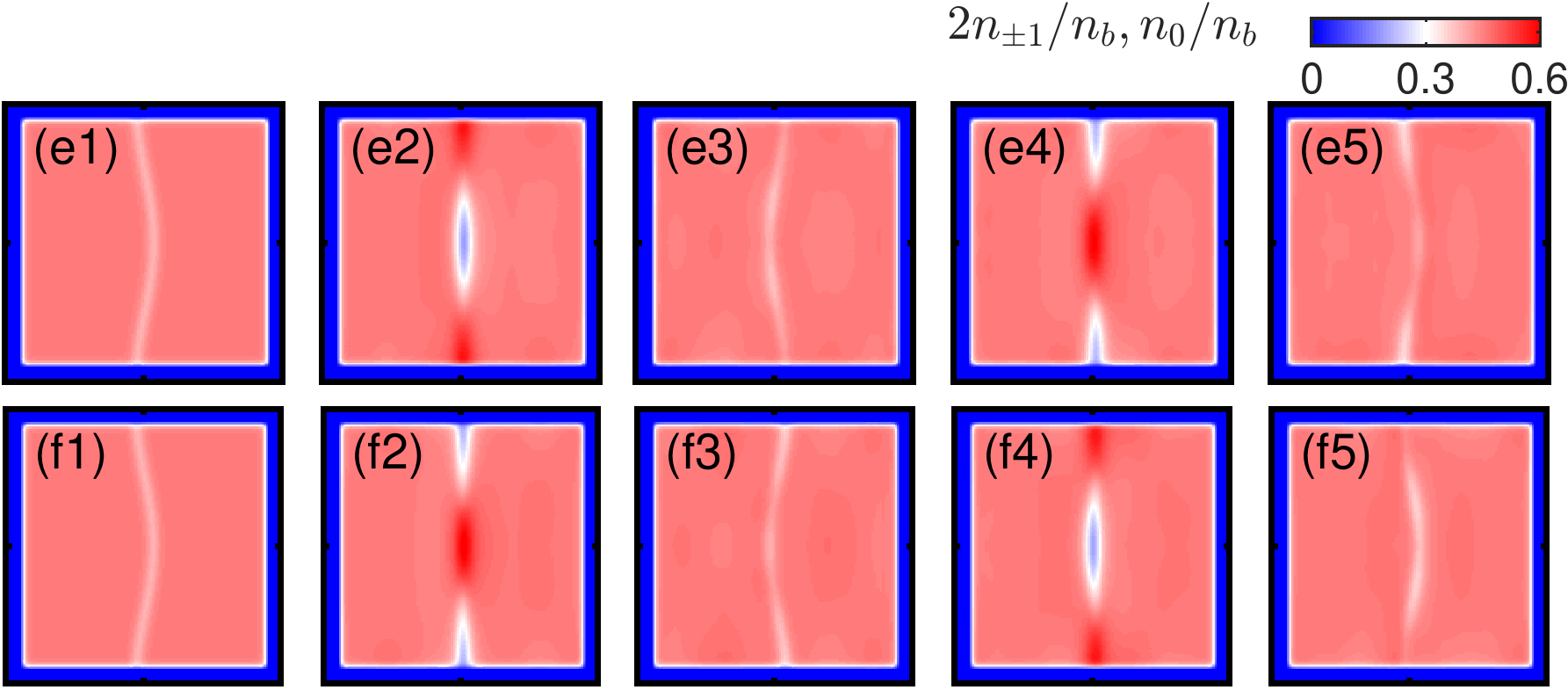}
	\caption{Spin-exchange dynamics during the MDW oscillation shown in  Fig.~\ref{f:realtimedynamics}(a1)-(a5).  (e1)-(e5), (f1)-f(5) show dynamics of component densities $2n_{\pm1}$ and $n_0$, respectively.
		\label{f:spinexchange}}
\end{figure}

It is worthwhile to mention that along with the domain wall oscillation, components of the wavefunction  exhibit spin-exchange dynamics [Fig.\ref{f:spinexchange}].

\begin{thebibliography}{62}%
	\makeatletter
	\providecommand \@ifxundefined [1]{%
		\@ifx{#1\undefined}
	}%
	\providecommand \@ifnum [1]{%
		\ifnum #1\expandafter \@firstoftwo
		\else \expandafter \@secondoftwo
		\fi
	}%
	\providecommand \@ifx [1]{%
		\ifx #1\expandafter \@firstoftwo
		\else \expandafter \@secondoftwo
		\fi
	}%
	\providecommand \natexlab [1]{#1}%
	\providecommand \enquote  [1]{``#1''}%
	\providecommand \bibnamefont  [1]{#1}%
	\providecommand \bibfnamefont [1]{#1}%
	\providecommand \citenamefont [1]{#1}%
	\providecommand \href@noop [0]{\@secondoftwo}%
	\providecommand \href [0]{\begingroup \@sanitize@url \@href}%
	\providecommand \@href[1]{\@@startlink{#1}\@@href}%
	\providecommand \@@href[1]{\endgroup#1\@@endlink}%
	\providecommand \@sanitize@url [0]{\catcode `\\12\catcode `\$12\catcode
		`\&12\catcode `\#12\catcode `\^12\catcode `\_12\catcode `\%12\relax}%
	\providecommand \@@startlink[1]{}%
	\providecommand \@@endlink[0]{}%
	\providecommand \url  [0]{\begingroup\@sanitize@url \@url }%
	\providecommand \@url [1]{\endgroup\@href {#1}{\urlprefix }}%
	\providecommand \urlprefix  [0]{URL }%
	\providecommand \Eprint [0]{\href }%
	\providecommand \doibase [0]{http://dx.doi.org/}%
	\providecommand \selectlanguage [0]{\@gobble}%
	\providecommand \bibinfo  [0]{\@secondoftwo}%
	\providecommand \bibfield  [0]{\@secondoftwo}%
	\providecommand \translation [1]{[#1]}%
	\providecommand \BibitemOpen [0]{}%
	\providecommand \bibitemStop [0]{}%
	\providecommand \bibitemNoStop [0]{.\EOS\space}%
	\providecommand \EOS [0]{\spacefactor3000\relax}%
	\providecommand \BibitemShut  [1]{\csname bibitem#1\endcsname}%
	\let\auto@bib@innerbib\@empty
	\bibitem [{\citenamefont {Bunkov}\ and\ \citenamefont
		{Godfrin}(2012)}]{bunkov2012topological}%
	\BibitemOpen
	\bibfield  {author} {\bibinfo {author} {\bibfnamefont {Yuriy~M}\ \bibnamefont
			{Bunkov}}\ and\ \bibinfo {author} {\bibfnamefont {Henri}\ \bibnamefont
			{Godfrin}},\ }\href@noop {} {\emph {\bibinfo {title} {Topological defects and
				the non-equilibrium dynamics of symmetry breaking phase transitions}}},\
	Vol.\ \bibinfo {volume} {549}\ (\bibinfo  {publisher} {Springer Science \&
		Business Media},\ \bibinfo {year} {2012})\BibitemShut {NoStop}%
	\bibitem [{\citenamefont {Manton}\ and\ \citenamefont
		{Sutcliffe}(2004)}]{manton2004topological}%
	\BibitemOpen
	\bibfield  {author} {\bibinfo {author} {\bibfnamefont {Nicholas}\
			\bibnamefont {Manton}}\ and\ \bibinfo {author} {\bibfnamefont {Paul}\
			\bibnamefont {Sutcliffe}},\ }\href@noop {} {\emph {\bibinfo {title}
			{Topological solitons}}}\ (\bibinfo  {publisher} {Cambridge University
		Press},\ \bibinfo {year} {2004})\BibitemShut {NoStop}%
	\bibitem [{\citenamefont {Eto}\ \emph {et~al.}(2014)\citenamefont {Eto},
		\citenamefont {Hirono}, \citenamefont {Nitta},\ and\ \citenamefont
		{Yasui}}]{eto2014vortices}%
	\BibitemOpen
	\bibfield  {author} {\bibinfo {author} {\bibfnamefont {Minoru}\ \bibnamefont
			{Eto}}, \bibinfo {author} {\bibfnamefont {Yuji}\ \bibnamefont {Hirono}},
		\bibinfo {author} {\bibfnamefont {Muneto}\ \bibnamefont {Nitta}}, \ and\
		\bibinfo {author} {\bibfnamefont {Shigehiro}\ \bibnamefont {Yasui}},\
	}\bibfield  {title} {\enquote {\bibinfo {title} {Vortices and other
				topological solitons in dense quark matter},}\ }\href@noop {} {\bibfield
		{journal} {\bibinfo  {journal} {Progress of Theoretical and Experimental
				Physics}\ }\textbf {\bibinfo {volume} {2014}} (\bibinfo {year}
		{2014})}\BibitemShut {NoStop}%
	\bibitem [{\citenamefont {Vilenkin}\ and\ \citenamefont
		{Shellard}(2000)}]{vilenkin2000cosmic}%
	\BibitemOpen
	\bibfield  {author} {\bibinfo {author} {\bibfnamefont {Alexander}\
			\bibnamefont {Vilenkin}}\ and\ \bibinfo {author} {\bibfnamefont {E~Paul~S}\
			\bibnamefont {Shellard}},\ }\href@noop {} {\emph {\bibinfo {title} {Cosmic
				strings and other topological defects}}}\ (\bibinfo  {publisher} {Cambridge
		University Press},\ \bibinfo {year} {2000})\BibitemShut {NoStop}%
	\bibitem [{\citenamefont {Kuznetsov}\ and\ \citenamefont
		{Turitsyn}(1988)}]{kuznetsov1988instability}%
	\BibitemOpen
	\bibfield  {author} {\bibinfo {author} {\bibfnamefont {EA}~\bibnamefont
			{Kuznetsov}}\ and\ \bibinfo {author} {\bibfnamefont {SK}~\bibnamefont
			{Turitsyn}},\ }\bibfield  {title} {\enquote {\bibinfo {title} {Instability
				and collapse of solitons in media with a defocusing nonlinearity},}\
	}\href@noop {} {\bibfield  {journal} {\bibinfo  {journal} {Zh. Eksp. Teor.
				Fiz}\ }\textbf {\bibinfo {volume} {94}},\ \bibinfo {pages} {129} (\bibinfo
		{year} {1988})}\BibitemShut {NoStop}%
	\bibitem [{\citenamefont {Muryshev}\ \emph {et~al.}(1999)\citenamefont
		{Muryshev}, \citenamefont {van Linden van~den Heuvell},\ and\ \citenamefont
		{Shlyapnikov}}]{Shlyapnikov1999}%
	\BibitemOpen
	\bibfield  {author} {\bibinfo {author} {\bibfnamefont {A.~E.}\ \bibnamefont
			{Muryshev}}, \bibinfo {author} {\bibfnamefont {H.~B.}\ \bibnamefont {van
				Linden van~den Heuvell}}, \ and\ \bibinfo {author} {\bibfnamefont {G.~V.}\
			\bibnamefont {Shlyapnikov}},\ }\bibfield  {title} {\enquote {\bibinfo {title}
			{Stability of standing matter waves in a trap},}\ }\href {\doibase
		10.1103/PhysRevA.60.R2665} {\bibfield  {journal} {\bibinfo  {journal} {Phys.
				Rev. A}\ }\textbf {\bibinfo {volume} {60}},\ \bibinfo {pages} {R2665--R2668}
		(\bibinfo {year} {1999})}\BibitemShut {NoStop}%
	\bibitem [{\citenamefont {Anderson}\ \emph {et~al.}(2001)\citenamefont
		{Anderson}, \citenamefont {Haljan}, \citenamefont {Regal}, \citenamefont
		{Feder}, \citenamefont {Collins}, \citenamefont {Clark},\ and\ \citenamefont
		{Cornell}}]{DSdecayExp}%
	\BibitemOpen
	\bibfield  {author} {\bibinfo {author} {\bibfnamefont {B.~P.}\ \bibnamefont
			{Anderson}}, \bibinfo {author} {\bibfnamefont {P.~C.}\ \bibnamefont
			{Haljan}}, \bibinfo {author} {\bibfnamefont {C.~A.}\ \bibnamefont {Regal}},
		\bibinfo {author} {\bibfnamefont {D.~L.}\ \bibnamefont {Feder}}, \bibinfo
		{author} {\bibfnamefont {L.~A.}\ \bibnamefont {Collins}}, \bibinfo {author}
		{\bibfnamefont {C.~W.}\ \bibnamefont {Clark}}, \ and\ \bibinfo {author}
		{\bibfnamefont {E.~A.}\ \bibnamefont {Cornell}},\ }\bibfield  {title}
	{\enquote {\bibinfo {title} {Watching dark solitons decay into vortex rings
				in a bose-einstein condensate},}\ }\href {\doibase
		10.1103/PhysRevLett.86.2926} {\bibfield  {journal} {\bibinfo  {journal}
			{Phys. Rev. Lett.}\ }\textbf {\bibinfo {volume} {86}},\ \bibinfo {pages}
		{2926--2929} (\bibinfo {year} {2001})}\BibitemShut {NoStop}%
	\bibitem [{\citenamefont {Huang}\ \emph {et~al.}(2003)\citenamefont {Huang},
		\citenamefont {Makarov},\ and\ \citenamefont {Velarde}}]{PhysRevA.67.023604}%
	\BibitemOpen
	\bibfield  {author} {\bibinfo {author} {\bibfnamefont {Guoxiang}\
			\bibnamefont {Huang}}, \bibinfo {author} {\bibfnamefont {Valeri~A.}\
			\bibnamefont {Makarov}}, \ and\ \bibinfo {author} {\bibfnamefont {Manuel~G.}\
			\bibnamefont {Velarde}},\ }\bibfield  {title} {\enquote {\bibinfo {title}
			{Two-dimensional solitons in bose-einstein condensates with a disk-shaped
				trap},}\ }\href {\doibase 10.1103/PhysRevA.67.023604} {\bibfield  {journal}
		{\bibinfo  {journal} {Phys. Rev. A}\ }\textbf {\bibinfo {volume} {67}},\
		\bibinfo {pages} {023604} (\bibinfo {year} {2003})}\BibitemShut {NoStop}%
	\bibitem [{\citenamefont {Son}\ and\ \citenamefont
		{Stephanov}(2002)}]{Son2002}%
	\BibitemOpen
	\bibfield  {author} {\bibinfo {author} {\bibfnamefont {D.~T.}\ \bibnamefont
			{Son}}\ and\ \bibinfo {author} {\bibfnamefont {M.~A.}\ \bibnamefont
			{Stephanov}},\ }\bibfield  {title} {\enquote {\bibinfo {title} {Domain walls
				of relative phase in two-component bose-einstein condensates},}\ }\href
	{\doibase 10.1103/PhysRevA.65.063621} {\bibfield  {journal} {\bibinfo
			{journal} {Phys. Rev. A}\ }\textbf {\bibinfo {volume} {65}},\ \bibinfo
		{pages} {063621} (\bibinfo {year} {2002})}\BibitemShut {NoStop}%
	\bibitem [{\citenamefont {Ihara}\ and\ \citenamefont
		{Kasamatsu}(2019)}]{Kasamatsu2019}%
	\BibitemOpen
	\bibfield  {author} {\bibinfo {author} {\bibfnamefont {Kousuke}\ \bibnamefont
			{Ihara}}\ and\ \bibinfo {author} {\bibfnamefont {Kenichi}\ \bibnamefont
			{Kasamatsu}},\ }\bibfield  {title} {\enquote {\bibinfo {title} {Transverse
				instability and disintegration of a domain wall of a relative phase in
				coherently coupled two-component bose-einstein condensates},}\ }\href
	{\doibase 10.1103/PhysRevA.100.013630} {\bibfield  {journal} {\bibinfo
			{journal} {Phys. Rev. A}\ }\textbf {\bibinfo {volume} {100}},\ \bibinfo
		{pages} {013630} (\bibinfo {year} {2019})}\BibitemShut {NoStop}%
	\bibitem [{\citenamefont {Gallem\'{\i}}\ \emph {et~al.}(2019)\citenamefont
		{Gallem\'{\i}}, \citenamefont {Pitaevskii}, \citenamefont {Stringari},\ and\
		\citenamefont {Recati}}]{Gallemi2019}%
	\BibitemOpen
	\bibfield  {author} {\bibinfo {author} {\bibfnamefont {A.}~\bibnamefont
			{Gallem\'{\i}}}, \bibinfo {author} {\bibfnamefont {L.~P.}\ \bibnamefont
			{Pitaevskii}}, \bibinfo {author} {\bibfnamefont {S.}~\bibnamefont
			{Stringari}}, \ and\ \bibinfo {author} {\bibfnamefont {A.}~\bibnamefont
			{Recati}},\ }\bibfield  {title} {\enquote {\bibinfo {title} {Decay of the
				relative phase domain wall into confined vortex pairs: The case of a
				coherently coupled bosonic mixture},}\ }\href {\doibase
		10.1103/PhysRevA.100.023607} {\bibfield  {journal} {\bibinfo  {journal}
			{Phys. Rev. A}\ }\textbf {\bibinfo {volume} {100}},\ \bibinfo {pages}
		{023607} (\bibinfo {year} {2019})}\BibitemShut {NoStop}%
	\bibitem [{\citenamefont {Qu}\ \emph {et~al.}(2016)\citenamefont {Qu},
		\citenamefont {Pitaevskii},\ and\ \citenamefont {Stringari}}]{MDQu2016}%
	\BibitemOpen
	\bibfield  {author} {\bibinfo {author} {\bibfnamefont {Chunlei}\ \bibnamefont
			{Qu}}, \bibinfo {author} {\bibfnamefont {Lev~P.}\ \bibnamefont {Pitaevskii}},
		\ and\ \bibinfo {author} {\bibfnamefont {Sandro}\ \bibnamefont {Stringari}},\
	}\bibfield  {title} {\enquote {\bibinfo {title} {Magnetic solitons in a
				binary bose-einstein condensate},}\ }\href {\doibase
		10.1103/PhysRevLett.116.160402} {\bibfield  {journal} {\bibinfo  {journal}
			{Phys. Rev. Lett.}\ }\textbf {\bibinfo {volume} {116}},\ \bibinfo {pages}
		{160402} (\bibinfo {year} {2016})}\BibitemShut {NoStop}%
	\bibitem [{\citenamefont {Farolfi}\ \emph {et~al.}(2020)\citenamefont
		{Farolfi}, \citenamefont {Trypogeorgos}, \citenamefont {Mordini},
		\citenamefont {Lamporesi},\ and\ \citenamefont {Ferrari}}]{MSexp1}%
	\BibitemOpen
	\bibfield  {author} {\bibinfo {author} {\bibfnamefont {A.}~\bibnamefont
			{Farolfi}}, \bibinfo {author} {\bibfnamefont {D.}~\bibnamefont
			{Trypogeorgos}}, \bibinfo {author} {\bibfnamefont {C.}~\bibnamefont
			{Mordini}}, \bibinfo {author} {\bibfnamefont {G.}~\bibnamefont {Lamporesi}},
		\ and\ \bibinfo {author} {\bibfnamefont {G.}~\bibnamefont {Ferrari}},\
	}\bibfield  {title} {\enquote {\bibinfo {title} {Observation of magnetic
				solitons in two-component bose-einstein condensates},}\ }\href {\doibase
		10.1103/PhysRevLett.125.030401} {\bibfield  {journal} {\bibinfo  {journal}
			{Phys. Rev. Lett.}\ }\textbf {\bibinfo {volume} {125}},\ \bibinfo {pages}
		{030401} (\bibinfo {year} {2020})}\BibitemShut {NoStop}%
	\bibitem [{\citenamefont {Chai}\ \emph {et~al.}(2020)\citenamefont {Chai},
		\citenamefont {Lao}, \citenamefont {Fujimoto}, \citenamefont {Hamazaki},
		\citenamefont {Ueda},\ and\ \citenamefont {Raman}}]{MSexp2}%
	\BibitemOpen
	\bibfield  {author} {\bibinfo {author} {\bibfnamefont {X.}~\bibnamefont
			{Chai}}, \bibinfo {author} {\bibfnamefont {D.}~\bibnamefont {Lao}}, \bibinfo
		{author} {\bibfnamefont {Kazuya}\ \bibnamefont {Fujimoto}}, \bibinfo {author}
		{\bibfnamefont {Ryusuke}\ \bibnamefont {Hamazaki}}, \bibinfo {author}
		{\bibfnamefont {Masahito}\ \bibnamefont {Ueda}}, \ and\ \bibinfo {author}
		{\bibfnamefont {C.}~\bibnamefont {Raman}},\ }\bibfield  {title} {\enquote
		{\bibinfo {title} {Magnetic solitons in a spin-1 bose-einstein condensate},}\
	}\href {\doibase 10.1103/PhysRevLett.125.030402} {\bibfield  {journal}
		{\bibinfo  {journal} {Phys. Rev. Lett.}\ }\textbf {\bibinfo {volume} {125}},\
		\bibinfo {pages} {030402} (\bibinfo {year} {2020})}\BibitemShut {NoStop}%
	\bibitem [{\citenamefont {Kang}\ \emph {et~al.}(2019)\citenamefont {Kang},
		\citenamefont {Seo}, \citenamefont {Takeuchi},\ and\ \citenamefont
		{Shin}}]{shin2019}%
	\BibitemOpen
	\bibfield  {author} {\bibinfo {author} {\bibfnamefont {Seji}\ \bibnamefont
			{Kang}}, \bibinfo {author} {\bibfnamefont {Sang~Won}\ \bibnamefont {Seo}},
		\bibinfo {author} {\bibfnamefont {Hiromitsu}\ \bibnamefont {Takeuchi}}, \
		and\ \bibinfo {author} {\bibfnamefont {Y.}~\bibnamefont {Shin}},\ }\bibfield
	{title} {\enquote {\bibinfo {title} {Observation of wall-vortex composite
				defects in a spinor bose-einstein condensate},}\ }\href {\doibase
		10.1103/PhysRevLett.122.095301} {\bibfield  {journal} {\bibinfo  {journal}
			{Phys. Rev. Lett.}\ }\textbf {\bibinfo {volume} {122}},\ \bibinfo {pages}
		{095301} (\bibinfo {year} {2019})}\BibitemShut {NoStop}%
	\bibitem [{\citenamefont {Ho}(1998)}]{Ho98}%
	\BibitemOpen
	\bibfield  {author} {\bibinfo {author} {\bibfnamefont {Tin-Lun}\ \bibnamefont
			{Ho}},\ }\bibfield  {title} {\enquote {\bibinfo {title} {Spinor bose
				condensates in optical traps},}\ }\href {\doibase 10.1103/PhysRevLett.81.742}
	{\bibfield  {journal} {\bibinfo  {journal} {Phys. Rev. Lett.}\ }\textbf
		{\bibinfo {volume} {81}},\ \bibinfo {pages} {742--745} (\bibinfo {year}
		{1998})}\BibitemShut {NoStop}%
	\bibitem [{\citenamefont {Ohmi}\ and\ \citenamefont {Machida}(1998)}]{OM98}%
	\BibitemOpen
	\bibfield  {author} {\bibinfo {author} {\bibfnamefont {Tetsuo}\ \bibnamefont
			{Ohmi}}\ and\ \bibinfo {author} {\bibfnamefont {Kazushige}\ \bibnamefont
			{Machida}},\ }\bibfield  {title} {\enquote {\bibinfo {title} {Bose-einstein
				condensation with internal degrees of freedom in alkali atom gases},}\ }\href
	{\doibase 10.1143/JPSJ.67.1822} {\bibfield  {journal} {\bibinfo  {journal}
			{Journal of the Physical Society of Japan}\ }\textbf {\bibinfo {volume}
			{67}},\ \bibinfo {pages} {1822--1825} (\bibinfo {year} {1998})}\BibitemShut
	{NoStop}%
	\bibitem [{\citenamefont {Sadler}\ \emph {et~al.}(2006)\citenamefont {Sadler},
		\citenamefont {Higbie}, \citenamefont {Leslie}, \citenamefont
		{Vengalattore},\ and\ \citenamefont {Stamper-Kurn}}]{Stampernatrue2006}%
	\BibitemOpen
	\bibfield  {author} {\bibinfo {author} {\bibfnamefont {L.~E.}\ \bibnamefont
			{Sadler}}, \bibinfo {author} {\bibfnamefont {J.~M.}\ \bibnamefont {Higbie}},
		\bibinfo {author} {\bibfnamefont {S.~R.}\ \bibnamefont {Leslie}}, \bibinfo
		{author} {\bibfnamefont {M.}~\bibnamefont {Vengalattore}}, \ and\ \bibinfo
		{author} {\bibfnamefont {D.~M.}\ \bibnamefont {Stamper-Kurn}},\ }\bibfield
	{title} {\enquote {\bibinfo {title} {Spontaneous symmetry breaking in a
				quenched ferromagnetic spinor bose–einstein condensate},}\ }\href
	{https://doi.org/10.1038/nature05094} {\bibfield  {journal} {\bibinfo
			{journal} {Nature}\ }\textbf {\bibinfo {volume} {443}},\ \bibinfo {pages}
		{312 EP --} (\bibinfo {year} {2006})}\BibitemShut {NoStop}%
	\bibitem [{\citenamefont {Stamper-Kurn}\ and\ \citenamefont
		{Ueda}(2013)}]{StamperRMP}%
	\BibitemOpen
	\bibfield  {author} {\bibinfo {author} {\bibfnamefont {Dan~M.}\ \bibnamefont
			{Stamper-Kurn}}\ and\ \bibinfo {author} {\bibfnamefont {Masahito}\
			\bibnamefont {Ueda}},\ }\bibfield  {title} {\enquote {\bibinfo {title}
			{Spinor bose gases: Symmetries, magnetism, and quantum dynamics},}\ }\href
	{\doibase 10.1103/RevModPhys.85.1191} {\bibfield  {journal} {\bibinfo
			{journal} {Rev. Mod. Phys.}\ }\textbf {\bibinfo {volume} {85}},\ \bibinfo
		{pages} {1191--1244} (\bibinfo {year} {2013})}\BibitemShut {NoStop}%
	\bibitem [{\citenamefont {Kawaguchi}\ and\ \citenamefont
		{Ueda}(2012)}]{KAWAGUCHI12}%
	\BibitemOpen
	\bibfield  {author} {\bibinfo {author} {\bibfnamefont {Yuki}\ \bibnamefont
			{Kawaguchi}}\ and\ \bibinfo {author} {\bibfnamefont {Masahito}\ \bibnamefont
			{Ueda}},\ }\bibfield  {title} {\enquote {\bibinfo {title} {Spinor
				bose–einstein condensates},}\ }\href {\doibase
		https://doi.org/10.1016/j.physrep.2012.07.005} {\bibfield  {journal}
		{\bibinfo  {journal} {Physics Reports}\ }\textbf {\bibinfo {volume} {520}},\
		\bibinfo {pages} {253 -- 381} (\bibinfo {year} {2012})},\ \bibinfo {note}
	{spinor Bose--Einstein condensates}\BibitemShut {NoStop}%
	\bibitem [{\citenamefont {Zhang}\ \emph {et~al.}(2005)\citenamefont {Zhang},
		\citenamefont {Zhou}, \citenamefont {Chang}, \citenamefont {Chapman},\ and\
		\citenamefont {You}}]{zhang2005DW}%
	\BibitemOpen
	\bibfield  {author} {\bibinfo {author} {\bibfnamefont {Wenxian}\ \bibnamefont
			{Zhang}}, \bibinfo {author} {\bibfnamefont {D.~L.}\ \bibnamefont {Zhou}},
		\bibinfo {author} {\bibfnamefont {M.-S.}\ \bibnamefont {Chang}}, \bibinfo
		{author} {\bibfnamefont {M.~S.}\ \bibnamefont {Chapman}}, \ and\ \bibinfo
		{author} {\bibfnamefont {L.}~\bibnamefont {You}},\ }\bibfield  {title}
	{\enquote {\bibinfo {title} {Dynamical instability and domain formation in a
				spin-1 bose-einstein condensate},}\ }\href {\doibase
		10.1103/PhysRevLett.95.180403} {\bibfield  {journal} {\bibinfo  {journal}
			{Phys. Rev. Lett.}\ }\textbf {\bibinfo {volume} {95}},\ \bibinfo {pages}
		{180403} (\bibinfo {year} {2005})}\BibitemShut {NoStop}%
	\bibitem [{\citenamefont {Higbie}\ \emph {et~al.}(2005)\citenamefont {Higbie},
		\citenamefont {Sadler}, \citenamefont {Inouye}, \citenamefont {Chikkatur},
		\citenamefont {Leslie}, \citenamefont {Moore}, \citenamefont {Savalli},\ and\
		\citenamefont {Stamper-Kurn}}]{Higbie2005}%
	\BibitemOpen
	\bibfield  {author} {\bibinfo {author} {\bibfnamefont {J.~M.}\ \bibnamefont
			{Higbie}}, \bibinfo {author} {\bibfnamefont {L.~E.}\ \bibnamefont {Sadler}},
		\bibinfo {author} {\bibfnamefont {S.}~\bibnamefont {Inouye}}, \bibinfo
		{author} {\bibfnamefont {A.~P.}\ \bibnamefont {Chikkatur}}, \bibinfo {author}
		{\bibfnamefont {S.~R.}\ \bibnamefont {Leslie}}, \bibinfo {author}
		{\bibfnamefont {K.~L.}\ \bibnamefont {Moore}}, \bibinfo {author}
		{\bibfnamefont {V.}~\bibnamefont {Savalli}}, \ and\ \bibinfo {author}
		{\bibfnamefont {D.~M.}\ \bibnamefont {Stamper-Kurn}},\ }\bibfield  {title}
	{\enquote {\bibinfo {title} {Direct nondestructive imaging of magnetization
				in a spin-1 bose-einstein gas},}\ }\href {\doibase
		10.1103/PhysRevLett.95.050401} {\bibfield  {journal} {\bibinfo  {journal}
			{Phys. Rev. Lett.}\ }\textbf {\bibinfo {volume} {95}},\ \bibinfo {pages}
		{050401} (\bibinfo {year} {2005})}\BibitemShut {NoStop}%
	\bibitem [{\citenamefont {Saito}\ and\ \citenamefont
		{Ueda}(2005)}]{Saito2005DW}%
	\BibitemOpen
	\bibfield  {author} {\bibinfo {author} {\bibfnamefont {Hiroki}\ \bibnamefont
			{Saito}}\ and\ \bibinfo {author} {\bibfnamefont {Masahito}\ \bibnamefont
			{Ueda}},\ }\bibfield  {title} {\enquote {\bibinfo {title} {Spontaneous
				magnetization and structure formation in a spin-1 ferromagnetic bose-einstein
				condensate},}\ }\href {\doibase 10.1103/PhysRevA.72.023610} {\bibfield
		{journal} {\bibinfo  {journal} {Phys. Rev. A}\ }\textbf {\bibinfo {volume}
			{72}},\ \bibinfo {pages} {023610} (\bibinfo {year} {2005})}\BibitemShut
	{NoStop}%
	\bibitem [{\citenamefont {Saito}\ \emph {et~al.}(2007)\citenamefont {Saito},
		\citenamefont {Kawaguchi},\ and\ \citenamefont
		{Ueda}}]{Saito2007pcvformation}%
	\BibitemOpen
	\bibfield  {author} {\bibinfo {author} {\bibfnamefont {Hiroki}\ \bibnamefont
			{Saito}}, \bibinfo {author} {\bibfnamefont {Yuki}\ \bibnamefont {Kawaguchi}},
		\ and\ \bibinfo {author} {\bibfnamefont {Masahito}\ \bibnamefont {Ueda}},\
	}\bibfield  {title} {\enquote {\bibinfo {title} {Topological defect formation
				in a quenched ferromagnetic bose-einstein condensates},}\ }\href {\doibase
		10.1103/PhysRevA.75.013621} {\bibfield  {journal} {\bibinfo  {journal} {Phys.
				Rev. A}\ }\textbf {\bibinfo {volume} {75}},\ \bibinfo {pages} {013621}
		(\bibinfo {year} {2007})}\BibitemShut {NoStop}%
	\bibitem [{\citenamefont {Zhang}\ \emph {et~al.}(2007)\citenamefont {Zhang},
		\citenamefont {M\"ustecapl\ifmmode \imath \else \i
			\fi{}o\ifmmode~\breve{g}\else \u{g}\fi{}lu},\ and\ \citenamefont
		{You}}]{zhang2007}%
	\BibitemOpen
	\bibfield  {author} {\bibinfo {author} {\bibfnamefont {Wenxian}\ \bibnamefont
			{Zhang}}, \bibinfo {author} {\bibfnamefont {\"O.~E.}\ \bibnamefont
			{M\"ustecapl\ifmmode \imath \else \i \fi{}o\ifmmode~\breve{g}\else
				\u{g}\fi{}lu}}, \ and\ \bibinfo {author} {\bibfnamefont {L.}~\bibnamefont
			{You}},\ }\bibfield  {title} {\enquote {\bibinfo {title} {Solitons in a
				trapped spin-1 atomic condensate},}\ }\href {\doibase
		10.1103/PhysRevA.75.043601} {\bibfield  {journal} {\bibinfo  {journal} {Phys.
				Rev. A}\ }\textbf {\bibinfo {volume} {75}},\ \bibinfo {pages} {043601}
		(\bibinfo {year} {2007})}\BibitemShut {NoStop}%
	\bibitem [{\citenamefont {Vengalattore}\ \emph {et~al.}(2008)\citenamefont
		{Vengalattore}, \citenamefont {Leslie}, \citenamefont {Guzman},\ and\
		\citenamefont {Stamper-Kurn}}]{Vengalattore2008a}%
	\BibitemOpen
	\bibfield  {author} {\bibinfo {author} {\bibfnamefont {M.}~\bibnamefont
			{Vengalattore}}, \bibinfo {author} {\bibfnamefont {S.~R.}\ \bibnamefont
			{Leslie}}, \bibinfo {author} {\bibfnamefont {J.}~\bibnamefont {Guzman}}, \
		and\ \bibinfo {author} {\bibfnamefont {D.~M.}\ \bibnamefont {Stamper-Kurn}},\
	}\bibfield  {title} {\enquote {\bibinfo {title} {Spontaneously modulated spin
				textures in a dipolar spinor bose-einstein condensate},}\ }\href {\doibase
		10.1103/PhysRevLett.100.170403} {\bibfield  {journal} {\bibinfo  {journal}
			{Phys. Rev. Lett.}\ }\textbf {\bibinfo {volume} {100}},\ \bibinfo {pages}
		{170403} (\bibinfo {year} {2008})}\BibitemShut {NoStop}%
	\bibitem [{\citenamefont {Kawaguchi}\ \emph {et~al.}(2010)\citenamefont
		{Kawaguchi}, \citenamefont {Saito}, \citenamefont {Kudo},\ and\ \citenamefont
		{Ueda}}]{Kawaguchi2010a}%
	\BibitemOpen
	\bibfield  {author} {\bibinfo {author} {\bibfnamefont {Yuki}\ \bibnamefont
			{Kawaguchi}}, \bibinfo {author} {\bibfnamefont {Hiroki}\ \bibnamefont
			{Saito}}, \bibinfo {author} {\bibfnamefont {Kazue}\ \bibnamefont {Kudo}}, \
		and\ \bibinfo {author} {\bibfnamefont {Masahito}\ \bibnamefont {Ueda}},\
	}\bibfield  {title} {\enquote {\bibinfo {title} {Spontaneous magnetic
				ordering in a ferromagnetic spinor dipolar bose-einstein condensate},}\
	}\href {\doibase 10.1103/PhysRevA.82.043627} {\bibfield  {journal} {\bibinfo
			{journal} {Phys. Rev. A}\ }\textbf {\bibinfo {volume} {82}},\ \bibinfo
		{pages} {043627} (\bibinfo {year} {2010})}\BibitemShut {NoStop}%
	\bibitem [{\citenamefont {Williamson}\ and\ \citenamefont
		{Blakie}(2016{\natexlab{a}})}]{Williamson2016a}%
	\BibitemOpen
	\bibfield  {author} {\bibinfo {author} {\bibfnamefont {Lewis~A.}\
			\bibnamefont {Williamson}}\ and\ \bibinfo {author} {\bibfnamefont {P.~B.}\
			\bibnamefont {Blakie}},\ }\bibfield  {title} {\enquote {\bibinfo {title}
			{Universal coarsening dynamics of a quenched ferromagnetic spin-1
				condensate},}\ }\href {\doibase 10.1103/PhysRevLett.116.025301} {\bibfield
		{journal} {\bibinfo  {journal} {Phys. Rev. Lett.}\ }\textbf {\bibinfo
			{volume} {116}},\ \bibinfo {pages} {025301} (\bibinfo {year}
		{2016}{\natexlab{a}})}\BibitemShut {NoStop}%
	\bibitem [{\citenamefont {Pr{\"u}fer}\ \emph {et~al.}(2018)\citenamefont
		{Pr{\"u}fer}, \citenamefont {Kunkel}, \citenamefont {Strobel}, \citenamefont
		{Lannig}, \citenamefont {Linnemann}, \citenamefont {Schmied}, \citenamefont
		{Berges}, \citenamefont {Gasenzer},\ and\ \citenamefont
		{Oberthaler}}]{Prufer2018a}%
	\BibitemOpen
	\bibfield  {author} {\bibinfo {author} {\bibfnamefont {Maximilian}\
			\bibnamefont {Pr{\"u}fer}}, \bibinfo {author} {\bibfnamefont {Philipp}\
			\bibnamefont {Kunkel}}, \bibinfo {author} {\bibfnamefont {Helmut}\
			\bibnamefont {Strobel}}, \bibinfo {author} {\bibfnamefont {Stefan}\
			\bibnamefont {Lannig}}, \bibinfo {author} {\bibfnamefont {Daniel}\
			\bibnamefont {Linnemann}}, \bibinfo {author} {\bibfnamefont
			{Christian-Marcel}\ \bibnamefont {Schmied}}, \bibinfo {author} {\bibfnamefont
			{J{\"u}rgen}\ \bibnamefont {Berges}}, \bibinfo {author} {\bibfnamefont
			{Thomas}\ \bibnamefont {Gasenzer}}, \ and\ \bibinfo {author} {\bibfnamefont
			{Markus~K.}\ \bibnamefont {Oberthaler}},\ }\bibfield  {title} {\enquote
		{\bibinfo {title} {Observation of universal dynamics in a spinor {Bose} gas
				far from equilibrium},}\ }\href {https://doi.org/10.1038/s41586-018-0659-0}
	{\bibfield  {journal} {\bibinfo  {journal} {Nature}\ }\textbf {\bibinfo
			{volume} {563}},\ \bibinfo {pages} {217--220} (\bibinfo {year}
		{2018})}\BibitemShut {NoStop}%
	\bibitem [{\citenamefont {Nistazakis}\ \emph {et~al.}(2008)\citenamefont
		{Nistazakis}, \citenamefont {Frantzeskakis}, \citenamefont {Kevrekidis},
		\citenamefont {Malomed},\ and\ \citenamefont
		{Carretero-Gonz\'alez}}]{nistazakis2008bright}%
	\BibitemOpen
	\bibfield  {author} {\bibinfo {author} {\bibfnamefont {H.~E.}\ \bibnamefont
			{Nistazakis}}, \bibinfo {author} {\bibfnamefont {D.~J.}\ \bibnamefont
			{Frantzeskakis}}, \bibinfo {author} {\bibfnamefont {P.~G.}\ \bibnamefont
			{Kevrekidis}}, \bibinfo {author} {\bibfnamefont {B.~A.}\ \bibnamefont
			{Malomed}}, \ and\ \bibinfo {author} {\bibfnamefont {R.}~\bibnamefont
			{Carretero-Gonz\'alez}},\ }\bibfield  {title} {\enquote {\bibinfo {title}
			{Bright-dark soliton complexes in spinor bose-einstein condensates},}\ }\href
	{\doibase 10.1103/PhysRevA.77.033612} {\bibfield  {journal} {\bibinfo
			{journal} {Phys. Rev. A}\ }\textbf {\bibinfo {volume} {77}},\ \bibinfo
		{pages} {033612} (\bibinfo {year} {2008})}\BibitemShut {NoStop}%
	\bibitem [{\citenamefont {Busch}\ and\ \citenamefont
		{Anglin}(2001)}]{Busch2001}%
	\BibitemOpen
	\bibfield  {author} {\bibinfo {author} {\bibfnamefont {Th.}\ \bibnamefont
			{Busch}}\ and\ \bibinfo {author} {\bibfnamefont {J.~R.}\ \bibnamefont
			{Anglin}},\ }\bibfield  {title} {\enquote {\bibinfo {title} {Dark-bright
				solitons in inhomogeneous bose-einstein condensates},}\ }\href {\doibase
		10.1103/PhysRevLett.87.010401} {\bibfield  {journal} {\bibinfo  {journal}
			{Phys. Rev. Lett.}\ }\textbf {\bibinfo {volume} {87}},\ \bibinfo {pages}
		{010401} (\bibinfo {year} {2001})}\BibitemShut {NoStop}%
	\bibitem [{\citenamefont {Bersano}\ \emph {et~al.}(2018)\citenamefont
		{Bersano}, \citenamefont {Gokhroo}, \citenamefont {Khamehchi}, \citenamefont
		{D'Ambroise}, \citenamefont {Frantzeskakis}, \citenamefont {Engels},\ and\
		\citenamefont {Kevrekidis}}]{ThreeComponentSoliton2018}%
	\BibitemOpen
	\bibfield  {author} {\bibinfo {author} {\bibfnamefont {T.~M.}\ \bibnamefont
			{Bersano}}, \bibinfo {author} {\bibfnamefont {V.}~\bibnamefont {Gokhroo}},
		\bibinfo {author} {\bibfnamefont {M.~A.}\ \bibnamefont {Khamehchi}}, \bibinfo
		{author} {\bibfnamefont {J.}~\bibnamefont {D'Ambroise}}, \bibinfo {author}
		{\bibfnamefont {D.~J.}\ \bibnamefont {Frantzeskakis}}, \bibinfo {author}
		{\bibfnamefont {P.}~\bibnamefont {Engels}}, \ and\ \bibinfo {author}
		{\bibfnamefont {P.~G.}\ \bibnamefont {Kevrekidis}},\ }\bibfield  {title}
	{\enquote {\bibinfo {title} {Three-component soliton states in spinor $f=1$
				bose-einstein condensates},}\ }\href {\doibase
		10.1103/PhysRevLett.120.063202} {\bibfield  {journal} {\bibinfo  {journal}
			{Phys. Rev. Lett.}\ }\textbf {\bibinfo {volume} {120}},\ \bibinfo {pages}
		{063202} (\bibinfo {year} {2018})}\BibitemShut {NoStop}%
	\bibitem [{\citenamefont {Manakov}(1974)}]{manakov1974theory}%
	\BibitemOpen
	\bibfield  {author} {\bibinfo {author} {\bibfnamefont {Sergei~V}\
			\bibnamefont {Manakov}},\ }\bibfield  {title} {\enquote {\bibinfo {title} {On
				the theory of two-dimensional stationary self-focusing of electromagnetic
				waves},}\ }\href@noop {} {\bibfield  {journal} {\bibinfo  {journal} {Soviet
				Physics-JETP}\ }\textbf {\bibinfo {volume} {38}},\ \bibinfo {pages}
		{248--253} (\bibinfo {year} {1974})}\BibitemShut {NoStop}%
	\bibitem [{foo({\natexlab{a}})}]{footnote-1}%
	\BibitemOpen
	\href@noop {} {\emph {\bibinfo {title} {\rm {The size of the system in
					$z$-direction is much smaller than the spin healing length
					$\xi_s=\hbar/\!\sqrt{2|g_s| M n_b}$.}}}}\BibitemShut {Stop}%
	\bibitem [{foo({\natexlab{b}})}]{footnote-3}%
	\BibitemOpen
	\href@noop {} {\emph {\bibinfo {title} {\rm{Generators of the rotational
					group $\textrm{SO}(3)$}: \bea S_x=\frac{1}{\sqrt{2}} \left(
				{\begin{array}{ccc} 0 & 1 & 0\\ 1 & 0 & 1\\ 0 & 1 & 0 \\ \end{array} }
				\right), S_y=\frac{i}{\sqrt{2}} \left( {\begin{array}{ccc} 0 & -1 & 0\\ 1 & 0
						& -1\\ 0 & 1 & 0 \\ \end{array} } \right), S_z=\left( {\begin{array}{ccc} 1 &
						0 & 0\\ 0 & 0 & 0\\ 0 & 0 & -1 \\ \end{array} } \right) \nn
				\eea}}}\BibitemShut {NoStop}%
	\bibitem [{mag()}]{magnetization}%
	\BibitemOpen
	\href@noop {} {\emph {\bibinfo {title}
			{\rm{$F_x=(\psi^{*}_0\psi_{-1}+\psi_0\psi^{*}_{+1}+\rm{h.c})/\sqrt{2}$;
					$F_y=[\psi_0^* (\psi_{-1}-\psi_{+1})-\rm{h.c}]/\sqrt{2}i$;
					$F_z=n_{+1}-n_{-1}$}}}}\BibitemShut {NoStop}%
	\bibitem [{foo({\natexlab{c}})}]{footnotescattering}%
	\BibitemOpen
	\href@noop {} {\emph {\bibinfo {title} {\rm{In terms of scattering length,
					the exactly solvable point corresponds to $4a_2=a_0$, where $a_F$ is the
					s-wave scattering length in the total spin $F$ channel. Here we used the
					following relation $g_n=4\pi \hbar^2(a_0+2a_2)/3M$ and $g_s=4\pi
					\hbar^2(a_2-a_0)/3M$~\cite{Ho98} }}}}\BibitemShut {NoStop}%
	\bibitem [{foo({\natexlab{d}})}]{footnoteSGS}%
	\BibitemOpen
	\href@noop {} {\emph {\bibinfo {title} {\rm{In BEC experiments, the drawback
					of introducing Rabi-coupling (magnetic noise) sometimes can be
					problematic~\cite{PhysRevA.97.013407, Ferrari2019}}}}}\BibitemShut {NoStop}%
	\bibitem [{\citenamefont {Ao}\ and\ \citenamefont {Chui}(1998)}]{Ao1998a}%
	\BibitemOpen
	\bibfield  {author} {\bibinfo {author} {\bibfnamefont {P.}~\bibnamefont
			{Ao}}\ and\ \bibinfo {author} {\bibfnamefont {S.~T.}\ \bibnamefont {Chui}},\
	}\bibfield  {title} {\enquote {\bibinfo {title} {Binary bose-einstein
				condensate mixtures in weakly and strongly segregated phases},}\ }\href
	{\doibase 10.1103/PhysRevA.58.4836} {\bibfield  {journal} {\bibinfo
			{journal} {Phys. Rev. A}\ }\textbf {\bibinfo {volume} {58}},\ \bibinfo
		{pages} {4836--4840} (\bibinfo {year} {1998})}\BibitemShut {NoStop}%
	\bibitem [{\citenamefont {Malomed}\ \emph {et~al.}(1990)\citenamefont
		{Malomed}, \citenamefont {Nepomnyashchy},\ and\ \citenamefont
		{Tribelsky}}]{Malomed1990}%
	\BibitemOpen
	\bibfield  {author} {\bibinfo {author} {\bibfnamefont {Boris~A.}\
			\bibnamefont {Malomed}}, \bibinfo {author} {\bibfnamefont {Alexander~A.}\
			\bibnamefont {Nepomnyashchy}}, \ and\ \bibinfo {author} {\bibfnamefont
			{Michael~I.}\ \bibnamefont {Tribelsky}},\ }\bibfield  {title} {\enquote
		{\bibinfo {title} {Domain boundaries in convection patterns},}\ }\href
	{\doibase 10.1103/PhysRevA.42.7244} {\bibfield  {journal} {\bibinfo
			{journal} {Phys. Rev. A}\ }\textbf {\bibinfo {volume} {42}},\ \bibinfo
		{pages} {7244--7263} (\bibinfo {year} {1990})}\BibitemShut {NoStop}%
	\bibitem [{foo({\natexlab{e}})}]{footnoteManakov}%
	\BibitemOpen
	\href@noop {} {\emph {\bibinfo {title} {\rm{Only in the Manakov limit, the
					two-component BEC possesses $\textrm{SU}(2)$ symmetry }}}}\BibitemShut
	{NoStop}%
	\bibitem [{\citenamefont {Chai}\ \emph {et~al.}(2021)\citenamefont {Chai},
		\citenamefont {Lao}, \citenamefont {Fujimoto},\ and\ \citenamefont
		{Raman}}]{chai2020magnetic}%
	\BibitemOpen
	\bibfield  {author} {\bibinfo {author} {\bibfnamefont {Xiao}\ \bibnamefont
			{Chai}}, \bibinfo {author} {\bibfnamefont {Di}~\bibnamefont {Lao}}, \bibinfo
		{author} {\bibfnamefont {Kazuya}\ \bibnamefont {Fujimoto}}, \ and\ \bibinfo
		{author} {\bibfnamefont {Chandra}\ \bibnamefont {Raman}},\ }\bibfield
	{title} {\enquote {\bibinfo {title} {Magnetic soliton: From two to three
				components with so(3) symmetry},}\ }\href {\doibase
		10.1103/PhysRevResearch.3.L012003} {\bibfield  {journal} {\bibinfo  {journal}
			{Phys. Rev. Research}\ }\textbf {\bibinfo {volume} {3}},\ \bibinfo {pages}
		{L012003} (\bibinfo {year} {2021})}\BibitemShut {NoStop}%
	\bibitem [{\citenamefont {Lim}\ and\ \citenamefont {Bao}(2008)}]{Lim2008a}%
	\BibitemOpen
	\bibfield  {author} {\bibinfo {author} {\bibfnamefont {Fong~Yin}\
			\bibnamefont {Lim}}\ and\ \bibinfo {author} {\bibfnamefont {Weizhu}\
			\bibnamefont {Bao}},\ }\bibfield  {title} {\enquote {\bibinfo {title}
			{Numerical methods for computing the ground state of spin-1 bose-einstein
				condensates in a uniform magnetic field},}\ }\href {\doibase
		10.1103/PhysRevE.78.066704} {\bibfield  {journal} {\bibinfo  {journal} {Phys.
				Rev. E}\ }\textbf {\bibinfo {volume} {78}},\ \bibinfo {pages} {066704}
		(\bibinfo {year} {2008})}\BibitemShut {NoStop}%
	\bibitem [{\citenamefont {Bao}\ and\ \citenamefont {Lim}(2008)}]{Bao2008a}%
	\BibitemOpen
	\bibfield  {author} {\bibinfo {author} {\bibfnamefont {Weizhu}\ \bibnamefont
			{Bao}}\ and\ \bibinfo {author} {\bibfnamefont {Fong~Yin}\ \bibnamefont
			{Lim}},\ }\bibfield  {title} {\enquote {\bibinfo {title} {Computing ground
				states of spin-1 bose–einstein condensates by the normalized gradient
				flow},}\ }\href {\doibase 10.1137/070698488} {\bibfield  {journal} {\bibinfo
			{journal} {SIAM Journal on Scientific Computing}\ }\textbf {\bibinfo {volume}
			{30}},\ \bibinfo {pages} {1925--1948} (\bibinfo {year} {2008})}\BibitemShut
	{NoStop}%
	\bibitem [{foo({\natexlab{f}})}]{footnotecurrent}%
	\BibitemOpen
	\href@noop {} {\emph {\bibinfo {title} {\rm{The details will be discussed
					elsewhere.}}}}\BibitemShut {Stop}%
	\bibitem [{\citenamefont {Barone}\ and\ \citenamefont
		{Paterno}(1982)}]{barone1982physics}%
	\BibitemOpen
	\bibfield  {author} {\bibinfo {author} {\bibfnamefont {Antonio}\ \bibnamefont
			{Barone}}\ and\ \bibinfo {author} {\bibfnamefont {Gianfranco}\ \bibnamefont
			{Paterno}},\ }\href@noop {} {\emph {\bibinfo {title} {Physics and
				applications of the Josephson effect}}}\ (\bibinfo  {publisher} {Wiley},\
	\bibinfo {year} {1982})\BibitemShut {NoStop}%
	\bibitem [{foo({\natexlab{g}})}]{footnotefreeends}%
	\BibitemOpen
	\href@noop {} {\emph {\bibinfo {title} {\rm{ Configurations with free ends
					are unstable }}}}\BibitemShut {NoStop}%
	\bibitem [{foo({\natexlab{h}})}]{footnotelargedeformation}%
	\BibitemOpen
	\href@noop {} {\emph {\bibinfo {title} {\rm{For large deformations of the
					magnetic domain wall from its equilibrium position the motion is periodic but
					is not harmonic }}}}\BibitemShut {NoStop}%
	\bibitem [{\citenamefont {Symes}\ \emph {et~al.}(2016)\citenamefont {Symes},
		\citenamefont {McLachlan},\ and\ \citenamefont {Blakie}}]{Symes2016a}%
	\BibitemOpen
	\bibfield  {author} {\bibinfo {author} {\bibfnamefont {L.~M.}\ \bibnamefont
			{Symes}}, \bibinfo {author} {\bibfnamefont {R.~I.}\ \bibnamefont
			{McLachlan}}, \ and\ \bibinfo {author} {\bibfnamefont {P.~B.}\ \bibnamefont
			{Blakie}},\ }\bibfield  {title} {\enquote {\bibinfo {title} {Efficient and
				accurate methods for solving the time-dependent spin-1 {Gross-Pitaevskii}
				equation},}\ }\href {\doibase 10.1103/PhysRevE.93.053309} {\bibfield
		{journal} {\bibinfo  {journal} {Phys. Rev. E}\ }\textbf {\bibinfo {volume}
			{93}},\ \bibinfo {pages} {053309} (\bibinfo {year} {2016})}\BibitemShut
	{NoStop}%
	\bibitem [{foo({\natexlab{i}})}]{footnoteBdG}%
	\BibitemOpen
	\href@noop {} {\emph {\bibinfo {title} {\rm{The BdG equation is $G
					(u,v)^{T}=E (u,v)^{T}$, where $G$ is the BdG operator. The symmetry $\sigma_x
					G \sigma_x=-G^{*}$, with $\sigma_x$ the $x$ Pauli matrix, ensures that
					$(v^{*},u^{*})^{T}$} is an eigenstate with eigenvalue $-E^{*}$. For the
				unstable modes, $\mathrm{Re}(E)$=0, hence $-E^{*}=E$. Then any linear
				combination of $(u,v)^{T}$ and $(v^{*},u^{*})^{T}$ is an eigenstate of $G$.
				Here the unstable modes are chosen to be real, in which case the two unstable
				modes are identical.}}}\BibitemShut {Stop}%
	\bibitem [{foo({\natexlab{j}})}]{footnote2}%
	\BibitemOpen
	\href@noop {} {\emph {\bibinfo {title} {\rm{The unstable modes must be
					localized in space and are insensitive to boundary conditions as long as
					$L_s\gg \ell$}}}}\BibitemShut {NoStop}%
	\bibitem [{foo({\natexlab{k}})}]{footnote3}%
	\BibitemOpen
	\href@noop {} {\emph {\bibinfo {title} {\rm{In 2D vector fields with three
					components, winding number is not well defined and hence the terminology
					vortex is quoted.}}}}\BibitemShut {Stop}%
	\bibitem [{\citenamefont {James}\ and\ \citenamefont
		{Lamacraft}(2011)}]{James2011}%
	\BibitemOpen
	\bibfield  {author} {\bibinfo {author} {\bibfnamefont {A.~J.~A.}\
			\bibnamefont {James}}\ and\ \bibinfo {author} {\bibfnamefont
			{A.}~\bibnamefont {Lamacraft}},\ }\bibfield  {title} {\enquote {\bibinfo
			{title} {Phase diagram of two-dimensional polar condensates in a magnetic
				field},}\ }\href {\doibase 10.1103/PhysRevLett.106.140402} {\bibfield
		{journal} {\bibinfo  {journal} {Phys. Rev. Lett.}\ }\textbf {\bibinfo
			{volume} {106}},\ \bibinfo {pages} {140402} (\bibinfo {year}
		{2011})}\BibitemShut {NoStop}%
	\bibitem [{\citenamefont {Kobayashi}(2019)}]{KobayashiBKT}%
	\BibitemOpen
	\bibfield  {author} {\bibinfo {author} {\bibfnamefont {Michikazu}\
			\bibnamefont {Kobayashi}},\ }\bibfield  {title} {\enquote {\bibinfo {title}
			{Berezinskii–kosterlitz–thouless transition of spin-1 spinor bose gases
				in the presence of the quadratic zeeman effect},}\ }\href {\doibase
		10.7566/JPSJ.88.094001} {\bibfield  {journal} {\bibinfo  {journal} {Journal
				of the Physical Society of Japan}\ }\textbf {\bibinfo {volume} {88}},\
		\bibinfo {pages} {094001} (\bibinfo {year} {2019})}\BibitemShut {NoStop}%
	\bibitem [{\citenamefont {Bourges}\ and\ \citenamefont
		{Blakie}(2017)}]{Andreane2017a}%
	\BibitemOpen
	\bibfield  {author} {\bibinfo {author} {\bibfnamefont {Andr\'eane}\
			\bibnamefont {Bourges}}\ and\ \bibinfo {author} {\bibfnamefont {P.~B.}\
			\bibnamefont {Blakie}},\ }\bibfield  {title} {\enquote {\bibinfo {title}
			{Different growth rates for spin and superfluid order in a quenched spinor
				condensate},}\ }\href {\doibase 10.1103/PhysRevA.95.023616} {\bibfield
		{journal} {\bibinfo  {journal} {Phys. Rev. A}\ }\textbf {\bibinfo {volume}
			{95}},\ \bibinfo {pages} {023616} (\bibinfo {year} {2017})}\BibitemShut
	{NoStop}%
	\bibitem [{\citenamefont {Williamson}\ and\ \citenamefont
		{Blakie}(2016{\natexlab{b}})}]{polarcorevortexLewis}%
	\BibitemOpen
	\bibfield  {author} {\bibinfo {author} {\bibfnamefont {Lewis~A.}\
			\bibnamefont {Williamson}}\ and\ \bibinfo {author} {\bibfnamefont {P.~B.}\
			\bibnamefont {Blakie}},\ }\bibfield  {title} {\enquote {\bibinfo {title}
			{Dynamics of polar-core spin vortices in a ferromagnetic spin-1 bose-einstein
				condensate},}\ }\href {\doibase 10.1103/PhysRevA.94.063615} {\bibfield
		{journal} {\bibinfo  {journal} {Phys. Rev. A}\ }\textbf {\bibinfo {volume}
			{94}},\ \bibinfo {pages} {063615} (\bibinfo {year}
		{2016}{\natexlab{b}})}\BibitemShut {NoStop}%
	\bibitem [{\citenamefont {Turner}(2009)}]{polarcorevortexTurner}%
	\BibitemOpen
	\bibfield  {author} {\bibinfo {author} {\bibfnamefont {Ari~M.}\ \bibnamefont
			{Turner}},\ }\bibfield  {title} {\enquote {\bibinfo {title} {Mass of a spin
				vortex in a bose-einstein condensate},}\ }\href {\doibase
		10.1103/PhysRevLett.103.080603} {\bibfield  {journal} {\bibinfo  {journal}
			{Phys. Rev. Lett.}\ }\textbf {\bibinfo {volume} {103}},\ \bibinfo {pages}
		{080603} (\bibinfo {year} {2009})}\BibitemShut {NoStop}%
	\bibitem [{\citenamefont {Chomaz}\ \emph {et~al.}(2015)\citenamefont {Chomaz},
		\citenamefont {Corman}, \citenamefont {Bienaim{\'e}}, \citenamefont
		{Desbuquois}, \citenamefont {Weitenberg}, \citenamefont {Nascimb{\`e}ne},
		\citenamefont {Beugnon},\ and\ \citenamefont {Dalibard}}]{Dalibard2015}%
	\BibitemOpen
	\bibfield  {author} {\bibinfo {author} {\bibfnamefont {Lauriane}\
			\bibnamefont {Chomaz}}, \bibinfo {author} {\bibfnamefont {Laura}\
			\bibnamefont {Corman}}, \bibinfo {author} {\bibfnamefont {Tom}\ \bibnamefont
			{Bienaim{\'e}}}, \bibinfo {author} {\bibfnamefont {R{\'e}mi}\ \bibnamefont
			{Desbuquois}}, \bibinfo {author} {\bibfnamefont {Christof}\ \bibnamefont
			{Weitenberg}}, \bibinfo {author} {\bibfnamefont {Sylvain}\ \bibnamefont
			{Nascimb{\`e}ne}}, \bibinfo {author} {\bibfnamefont {J{\'e}r{\^o}me}\
			\bibnamefont {Beugnon}}, \ and\ \bibinfo {author} {\bibfnamefont {Jean}\
			\bibnamefont {Dalibard}},\ }\bibfield  {title} {\enquote {\bibinfo {title}
			{Emergence of coherence via transverse condensation in a uniform
				quasi-two-dimensional bose gas},}\ }\href
	{http://www.nature.com/articles/ncomms7162} {\bibfield  {journal} {\bibinfo
			{journal} {Nature communications}\ }\textbf {\bibinfo {volume} {6}} (\bibinfo
		{year} {2015})}\BibitemShut {NoStop}%
	\bibitem [{\citenamefont {Gauthier}\ \emph {et~al.}(2016)\citenamefont
		{Gauthier}, \citenamefont {Lenton}, \citenamefont {Parry}, \citenamefont
		{Baker}, \citenamefont {Davis}, \citenamefont {Rubinsztein-Dunlop},\ and\
		\citenamefont {Neely}}]{Gauthier16}%
	\BibitemOpen
	\bibfield  {author} {\bibinfo {author} {\bibfnamefont {G.}~\bibnamefont
			{Gauthier}}, \bibinfo {author} {\bibfnamefont {I.}~\bibnamefont {Lenton}},
		\bibinfo {author} {\bibfnamefont {N.~McKay}\ \bibnamefont {Parry}}, \bibinfo
		{author} {\bibfnamefont {M.}~\bibnamefont {Baker}}, \bibinfo {author}
		{\bibfnamefont {M.~J.}\ \bibnamefont {Davis}}, \bibinfo {author}
		{\bibfnamefont {H.}~\bibnamefont {Rubinsztein-Dunlop}}, \ and\ \bibinfo
		{author} {\bibfnamefont {T.~W.}\ \bibnamefont {Neely}},\ }\bibfield  {title}
	{\enquote {\bibinfo {title} {Direct imaging of a digital-micromirror device
				for configurable microscopic optical potentials},}\ }\href {\doibase
		10.1364/OPTICA.3.001136} {\bibfield  {journal} {\bibinfo  {journal}
			{\color{blue}Optica}\ }\textbf {\bibinfo {volume} {3}},\ \bibinfo {pages}
		{1136--1143} (\bibinfo {year} {2016})}\BibitemShut {NoStop}%
	\bibitem [{\citenamefont {Huh}\ \emph {et~al.}(2020)\citenamefont {Huh},
		\citenamefont {Kim}, \citenamefont {Kwon},\ and\ \citenamefont
		{Choi}}]{Huh2020a}%
	\BibitemOpen
	\bibfield  {author} {\bibinfo {author} {\bibfnamefont {SeungJung}\
			\bibnamefont {Huh}}, \bibinfo {author} {\bibfnamefont {Kyungtae}\
			\bibnamefont {Kim}}, \bibinfo {author} {\bibfnamefont {Kiryang}\ \bibnamefont
			{Kwon}}, \ and\ \bibinfo {author} {\bibfnamefont {Jae-yoon}\ \bibnamefont
			{Choi}},\ }\bibfield  {title} {\enquote {\bibinfo {title} {Observation of a
				strongly ferromagnetic spinor bose-einstein condensate},}\ }\href {\doibase
		10.1103/PhysRevResearch.2.033471} {\bibfield  {journal} {\bibinfo  {journal}
			{Phys. Rev. Research}\ }\textbf {\bibinfo {volume} {2}},\ \bibinfo {pages}
		{033471} (\bibinfo {year} {2020})}\BibitemShut {NoStop}%
	\bibitem [{\citenamefont {Trypogeorgos}\ \emph {et~al.}(2018)\citenamefont
		{Trypogeorgos}, \citenamefont {Vald\'es-Curiel}, \citenamefont {Lundblad},\
		and\ \citenamefont {Spielman}}]{PhysRevA.97.013407}%
	\BibitemOpen
	\bibfield  {author} {\bibinfo {author} {\bibfnamefont {D.}~\bibnamefont
			{Trypogeorgos}}, \bibinfo {author} {\bibfnamefont {A.}~\bibnamefont
			{Vald\'es-Curiel}}, \bibinfo {author} {\bibfnamefont {N.}~\bibnamefont
			{Lundblad}}, \ and\ \bibinfo {author} {\bibfnamefont {I.~B.}\ \bibnamefont
			{Spielman}},\ }\bibfield  {title} {\enquote {\bibinfo {title} {Synthetic
				clock transitions via continuous dynamical decoupling},}\ }\href {\doibase
		10.1103/PhysRevA.97.013407} {\bibfield  {journal} {\bibinfo  {journal} {Phys.
				Rev. A}\ }\textbf {\bibinfo {volume} {97}},\ \bibinfo {pages} {013407}
		(\bibinfo {year} {2018})}\BibitemShut {NoStop}%
	\bibitem [{\citenamefont {Farolfi}\ \emph {et~al.}(2019)\citenamefont
		{Farolfi}, \citenamefont {Trypogeorgos}, \citenamefont {Colzi}, \citenamefont
		{Fava}, \citenamefont {Lamporesi},\ and\ \citenamefont
		{Ferrari}}]{Ferrari2019}%
	\BibitemOpen
	\bibfield  {author} {\bibinfo {author} {\bibfnamefont {A.}~\bibnamefont
			{Farolfi}}, \bibinfo {author} {\bibfnamefont {D.}~\bibnamefont
			{Trypogeorgos}}, \bibinfo {author} {\bibfnamefont {G.}~\bibnamefont {Colzi}},
		\bibinfo {author} {\bibfnamefont {E.}~\bibnamefont {Fava}}, \bibinfo {author}
		{\bibfnamefont {G.}~\bibnamefont {Lamporesi}}, \ and\ \bibinfo {author}
		{\bibfnamefont {G.}~\bibnamefont {Ferrari}},\ }\bibfield  {title} {\enquote
		{\bibinfo {title} {Design and characterization of a compact magnetic shield
				for ultracold atomic gas experiments},}\ }\href {\doibase 10.1063/1.5119915}
	{\bibfield  {journal} {\bibinfo  {journal} {Review of Scientific
				Instruments}\ }\textbf {\bibinfo {volume} {90}},\ \bibinfo {pages} {115114}
		(\bibinfo {year} {2019})}\BibitemShut {NoStop}%
\end{thebibliography}

%

\end{document}